\documentclass[final,5p,times,twocolumn,authoryear]{elsarticle}

\usepackage{amssymb}

\usepackage{amsmath}
\usepackage{hyperref}

\journal{New Astronomy}

\begin{document}

\begin{frontmatter}



\title{HD\,3191, the high-mass X-ray binary that wasn't there\tnoteref{t1}}
\tnotetext[t1]{Based on optical spectra collected with the TIGRE telescope (La Luz, Mexico).}


\author[1]{Gregor Rauw\corref{cor1}}
\ead{g.rauw@uliege.be} 
\author[1]{Ya\"el Naz\'e\fnref{fn1}}
\author[1,2,3]{Piotr Antoni Ko{\l}aczek-Szyma{\'n}ski}
\cortext[cor1]{Corresponding author}
\affiliation[1]{organization={Space sciences, Technologies and Astrophysics Research (STAR) Institute, Universit\'e de Li\`ege},
            addressline={All\'ee du 6 Ao\^ut, 19c, B\^at B5c}, 
            city={Li\`ege},
            postcode={4000}, 
            country={Belgium}}
\affiliation[2]{organization={Astronomical Observatory of the University of Warsaw},
            addressline={Al. Ujazdowskie 4}, 
            city={Warsaw},
            postcode={00-478}, 
            country={Poland}}
\affiliation[3]{organization={University of Wroc{\l}aw, Faculty of Physics and Astronomy, Astronomical Institute},
            addressline={ul. Kopernika 11}, 
            city={Wroc{\l}aw},
            postcode={51-622}, 
            country={Poland}}
\fntext[fn1]{Senior research associate FRS-FNRS (Belgium)}

\begin{abstract}
  The rapidly rotating B1\,IV:nn star HD\,3191 lies within the error box of a flaring {\it Fermi} $\gamma$-ray source. Although the counterpart of the {\it Fermi} source is likely an active galaxy, HD\,3191 has nevertheless been suggested to be a high-mass X-ray binary, possibly hosting a black hole companion. The star displays roughly sinusoidal photometric variations with amplitudes of $\sim 12$\,mmag for two frequencies $\nu_1 = 0.1235$\,d$^{-1}$ and $\nu_2 = 1.6038$\,d$^{-1}$. Half of the former frequency ($\nu_1/2$) had previously been interpreted as the orbital frequency of a high-mass X-ray binary in which the B1\,IV:nn primary undergoes ellipsoidal variations. We show that this scenario fails to account for the lack of significant radial velocity variations and for the overall properties of the star. Our spectroscopic observations instead unveil line profile variations, suggesting that the photometric and line profile variations arise from multi-mode pulsations rather than from orbital effects. Whilst we cannot rule out the possibility that HD\,3191 could be a nascent binary, consisting of a B1\,IV:nn with a low-mass pre-main sequence companion, the most likely scenario is a single star displaying non-radial pulsations. The fast stellar rotation would thus be intrinsic to the B1\,IV:nn star rather than being the result of a spin-up during a past mass-transfer episode.
\end{abstract}



\begin{keyword}
stars: early-type \sep stars: individual (HD\,3191) \sep binaries: close \sep stars: variables: general \sep stars: oscillations (including pulsations)


\end{keyword}

\end{frontmatter}



\section{Introduction}
Over the past two decades, the interest of the astrophysical community in the evolution of massive stars and massive binary systems enjoyed a strong boost. Observationally, it was found that the majority of massive stars form and evolve in binary or higher multiplicity systems \citep[e.g.][and references therein]{San12,Off23}. In the course of their evolution, such systems can go through a phase where the most massive component has already exploded as a supernova, leaving behind a neutron star or a black hole orbiting the (initially less massive) non-degenerate companion \citep[e.g.,][]{Kru18,Man24}. Such a configuration frequently leads to situations where the non-degenerate star transfers material and angular momentum to its compact companion. Such systems appear as high-mass X-ray binaries (HMXBs) that emit copious amounts of X-rays \citep{For23}. In some HMXBs, such as LS\,I~+61\,303, the high-energy emission extends into the $\gamma$-ray domain \citep[e.g.,][]{Alb08}.

Our target, HD\,3191, has been cited as a candidate of such a $\gamma$-ray binary. \citet{Piv16} reported the detection with the {\it Fermi} satellite of a previously unknown transient $\gamma$-ray source (Fermi J0035+6131) near the Galactic Plane. This source subsequently entered the {\it Fermi} catalog as 4FGL\,J0035.8+6131. Based on an {\it XMM-Newton} observation, \citet{Pan16,Pan18} reported that the brightest X-ray source inside the error region of 4FGL\,J0035.8+6131 coincides in position with the compact radio source 87GB\,003232.7+611352 (VCS4\,J0035+6130). This latter object is likely an active galaxy located behind the disk of the Milky Way, and thus appears to be the most likely counterpart of the transient $\gamma$-ray source. The second-brightest X-ray source in the field was associated with HD\,3191. \citet{Pan18} found that the X-ray spectrum of HD\,3191 could be equally well fitted with an absorbed bremsstrahlung (with $kT \geq 6$\,keV and $N_{\rm H} = (4.1 \pm 2.2) \times 10^{21}$\,cm$^{-2}$) or an absorbed power law model (photon index of $1.5 \pm 0.4$ and $N_{\rm H} = (4.5 \pm 3.0) \times 10^{21}$\,cm$^{-2}$). They inferred an observed ﬂux of $\sim 10^{-13}$\,erg\,cm$^{-2}$\,s$^{-1}$ in the 0.5 -- 12\,keV energy range. Because of the hardness of the X-ray spectrum, \citet{Pan16,Pan18} suggested HD\,3191 to be a HMXB with a comparatively low X-ray luminosity, although it was deemed unlikely to be the counterpart of 4FGL\,J0035.8+6131. 

An optical spectrum of HD\,3191, obtained by \citet{Mun16} in March 2016, revealed no peculiarity such as emission lines or spectroscopic signatures of a companion star. \citet{Mun16} measured a stellar heliocentric radial velocity (RV) of $-46.0 \pm 0.5$\,km\,s$^{-1}$, as well as a projected stellar rotational velocity of $v\,\sin{i} = 265 \pm 10$\,km\,s$^{-1}$. They pointed out that their RV value did not match the value of $-22 \pm 3$\,km\,s$^{-1}$ reported by \citet{Pet61}. \citet{Mun16} thus suggested that HD\,3191 could be a single-lined binary (SB1) system. \citet{Mar21} reported on a low resolution optical spectrum that apparently displayed a weak He\,{\sc ii} $\lambda$\,4686 emission (EW $\sim -0.54$\,\AA) on top of the typical spectrum of a B1\,IV:nn star. Ground-based photometric measurements made by \citet{Mar21} unveiled a period near 8\,days which they interpreted as ellipsoidal variations of the B1 star orbiting a compact companion every 16\,days. Two sectors of space-borne photometry (Sectors 17 and 18) collected with the Transiting Exoplanet Survey Satellite \citep[{\it TESS},][]{TESS} confirmed the presence of a $\sim 8$\,d periodicity along with another period of 0.623\,d. \citet{Mar21} suggested that the 0.623\,d period could reflect either rotational modulation or pulsations. In a follow-up study, \citet{Mar23} presented a set of RV measurements that they tentatively interpreted as orbital motion with a low amplitude in a HMXB, possibly hosting a black hole. An alternative interpretation of the earliest set of {\it TESS} data was proposed by \citet{Bal20} who classified HD\,3191 as a slowly pulsating B-type star (SPB).

In this work, we revisit the properties of HD\,3191. We have collected a series of new echelle spectra of HD\,3191 to search for putative binary signatures and take advantage of a more extensive set of {\it TESS} observations. Section\,\ref{sect:obs} presents the observations used in this paper. These data are analysed in Sect.\,\ref{sect:analysis} and their interpretation is discussed in Sect.\,\ref{sect:discussion}. Finally, Sect.\,\ref{sect:conclusion} briefly summarises our conclusions. 

\section{Observations and data processing \label{sect:obs}}
\subsection{Spectroscopic data \label{obs:spectro}}
Given the uncertainties on the nature of HD\,3191, we set up a spectroscopic monitoring campaign to find out whether the star undergoes orbital motion and to check whether the spectrum indeed exhibits He\,{\sc ii} $\lambda$~4686 emission. Thirty echelle spectra of HD\,3191 were collected between October 2021 and September 2023 with the refurbished HEROS echelle spectrograph \citep{Kaufer2,Schmitt} on the robotic 1.2\,m TIGRE telescope \citep{Schmitt,Gon22} at La Luz Observatory near Guanajuato (Mexico). The TIGRE/HEROS spectrograph has a resolving power of 20\,000 over the wavelength range from 3760 -- 8700\,\AA\ with a small gap around 5600\,\AA. The data were processed with the HEROS reduction pipeline \citep{Mittag,Schmitt}. Telluric absorption lines in the spectral regions around the He\,{\sc i} $\lambda$\,5876 and H$\alpha$ lines were removed by means of the {\tt telluric} tool of the {\sc iraf} software and using the atlas of telluric lines of \citet{Hinkle}. Further data analysis (continuum normalization, RV measurements) was performed with the {\sc midas} software. 

\subsection{Photometry}
Space-borne photometry of HD\,3191 was obtained in six sectors with the {\it TESS} satellite. {\it TESS} features four wide-field cameras operating in the 6000\,\AA\, to 1\,$\mu$m waveband. Each camera covers a $24^{\circ} \times 24^{\circ}$ field with pixel sizes of 15\,$\mu$m squared corresponding to (21\,arcsec)$^2$ on the sky. {\it TESS} monitors sky sectors of $24^{\circ} \times 96^{\circ}$ for about 27 consecutive days. HD\,3191 was observed at a 2\,min cadence during Sectors 17, 18 and 85, with a 30\,min cadence during Sector 24, and every 200\,s during Sectors 58 and 78.

The 2\,min cadence light curves, processed with the {\it TESS} pipeline \citep{Jen16}, were downloaded from the Mikulski Archive for Space Telescopes (MAST) portal\footnote{http://mast.stsci.edu/}. They provide simple background-corrected aperture photometry (SAP) as well as Pre-search Data Conditioned (PDC) photometry obtained after correcting trends correlated with systematic spacecraft or instrumental effects. In the present case, both SAP and PDC light curves were found to be very similar. However, since we found an apparently lower amplitude for the 8\,day modulation in some of the pipeline-produced photometry, we also analysed the full frame images (FFI) taken at a cadence of 30\,min (Sectors 17 and 18) or 200\,s (Sector 85). The FFI data were processed with the Lightkurve Python software package\footnote{https://docs.lightkurve.org/}. Aperture photometry was extracted on $51 \times 51$ pixels image cutouts. For the source mask, we adopted a ﬂux threshold of $20$ times the median absolute deviation over the median ﬂux\footnote{For Sector 17, a threshold of 30 times the median absolute deviation gave better results and was thus adopted.}. The background was evaluated either from those pixels in the cutout that were below the median ﬂux, or by means of a principal component analysis (pca) including five components. The background-corrected ﬂuxes were converted to magnitudes and data points with errors larger than the mean error augmented by three times the dispersion of the errors were removed. The same procedure was applied to the FFI data taken during Sectors 24 (30\,min cadence), 58 and 78 (both 200\,s cadence). The results of both background subtraction methods were generally in very good agreement, with the exception of the data from Sector 78 which displayed small differences between the pca and median fluxes method. We subsequently used the photometry obtained with the pca method, except for Sector 78 for which the pca lightcurve showed a weaker signature of the main modulation and where we used the median fluxes instead. 

Since the {\it TESS} photometry of a given source is extracted over several pixels, the light curves of objects in crowded regions can be contaminated by neighbouring sources. To check whether this is the case for HD\,3191, we queried the third {\it Gaia} data release catalog \citep[DR3,][]{DR3}. We found a total of 70 additional objects within a 1\,arcmin radius of HD\,3191. However, the brightest neighbouring source is 4.3\,mag fainter than HD\,3191 in the {\it Gaia} $G_{RP}$ band implying that the {\it TESS} photometry of HD\,3191 should be free of significant contamination. This conclusion is backed-up by the value of the {\it TESS} CROWDSAP parameter which estimates the fraction of (background corrected) flux in the photometric aperture attributable to the target. This keyword amounts to 0.978 for Sector 17, 0.968 for Sector 18, and 0.969 for Sector 85.

\subsection{X-ray data}
HD\,3191 was observed twice with the {\it XMM-Newton} satellite \citep{Jan01}. We retrieved these data from the archive and processed them with the Science Analysis System (SAS) software version 21.0.0 and the current calibration files available in December 2025. 

A first observation was obtained in February 2008 (JD\,2454506.501 at mid exposure). The field of view was centred on the HMXB IGR\,J00370+6122, and the observation was taken in small window mode and with the medium optical bloccking filter. HD\,3191 fell on the outer part of the field of view of the EPIC-MOS1 and MOS2 instruments, but was not covered by the pn detector. The observation was affected by several background flares that were filtered out, resulting in effective exposure times of 12.3 and 16.6\,ks respectively for MOS1 and MOS2. HD\,3191 remained undetected in this observation.

The second observation, with the aimpoint set on the {\it Fermi} source J0035+6131 and using the full frame observing mode along with the medium optical blocking filter, was taken in January 2016 (HJD\,2457411.749 at mid exposure). No background flare affected this observation, resulting in exposure times of 11.5 and 8.9\,ks respectively for both MOS instruments and the EPIC-pn camera. HD\,3191 fell into the inner part of the field of view and was detected as a rather faint source with all three EPIC cameras. The MOS2 extraction region was adjusted to exclude two bad CCD columns crossing the source region. The background-corrected count rates were $(5.04 \pm 0.89) \times 10^{-3}$\,cps, $(3.13 \pm 0.71) \times 10^{-3}$\,cps and $(13.13 \pm 1.87) \times 10^{-3}$\,cps respectively for the MOS1, MOS2 and pn detectors. Only the EPIC-pn spectrum was of sufficient quality to perform a spectral analysis.
\begin{figure}[h]
    \begin{center}
          \resizebox{8.5cm}{!}{\includegraphics{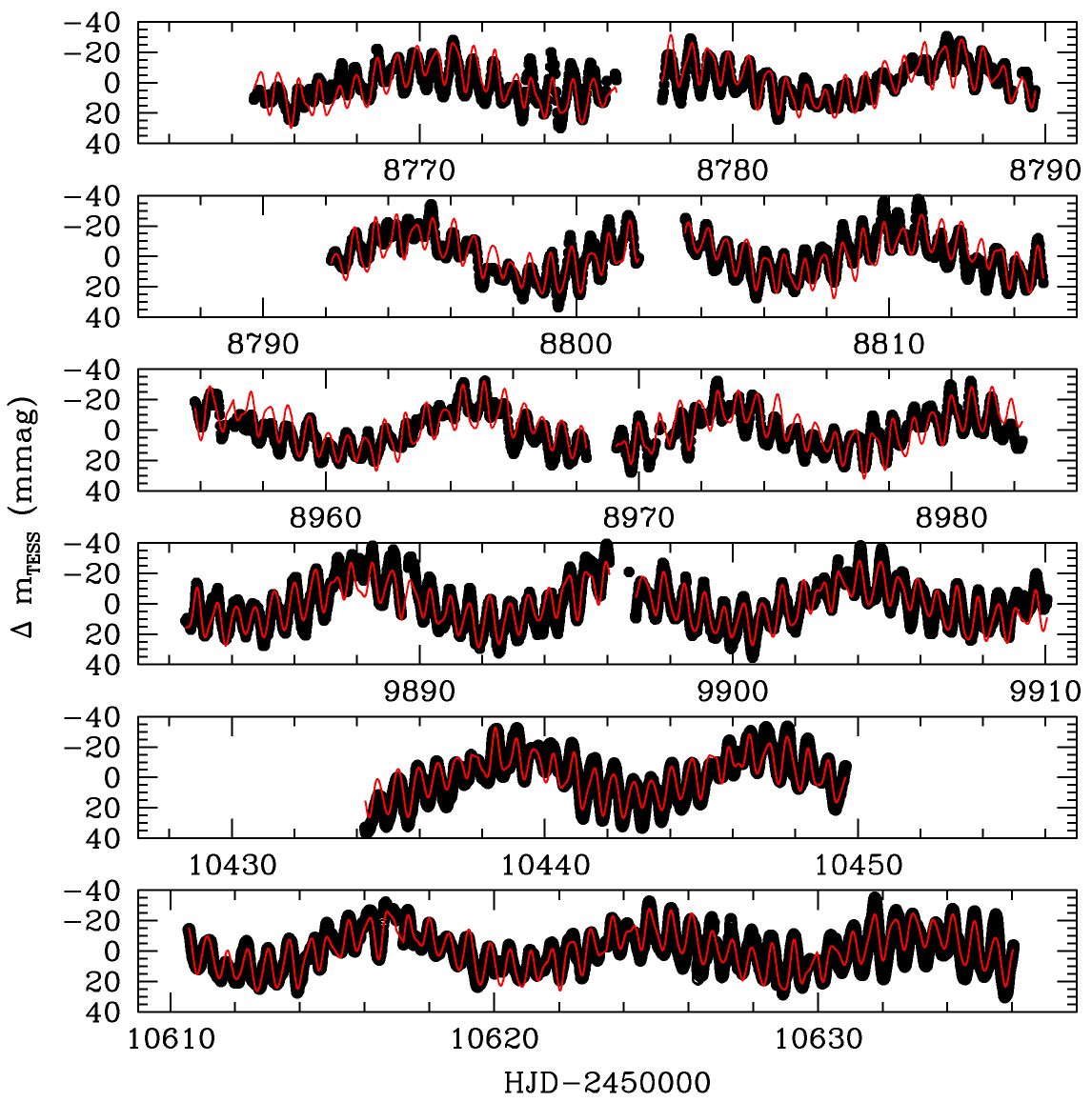}}
      \caption{{\it TESS} light curves of HD\,3191 as observed during Sectors 17, 18, 24, 58, 78 and 85 (from top to bottom). The photometry was extracted from the FFI data with a cadence of 30\,min (Sectors 17, 18 and 24) and 200\,s (Sectors 58, 78 and 85). The red curve yields the best-fit adjustment of variations with eight frequencies listed in Table\,\ref{tab_comb}.\label{TESSlc}}
    \end{center}
\end{figure}
\section{Results \label{sect:analysis}}
\subsection{Photometric variability}
\begin{figure}[h]
    \begin{center}
      \resizebox{8.5cm}{!}{\includegraphics{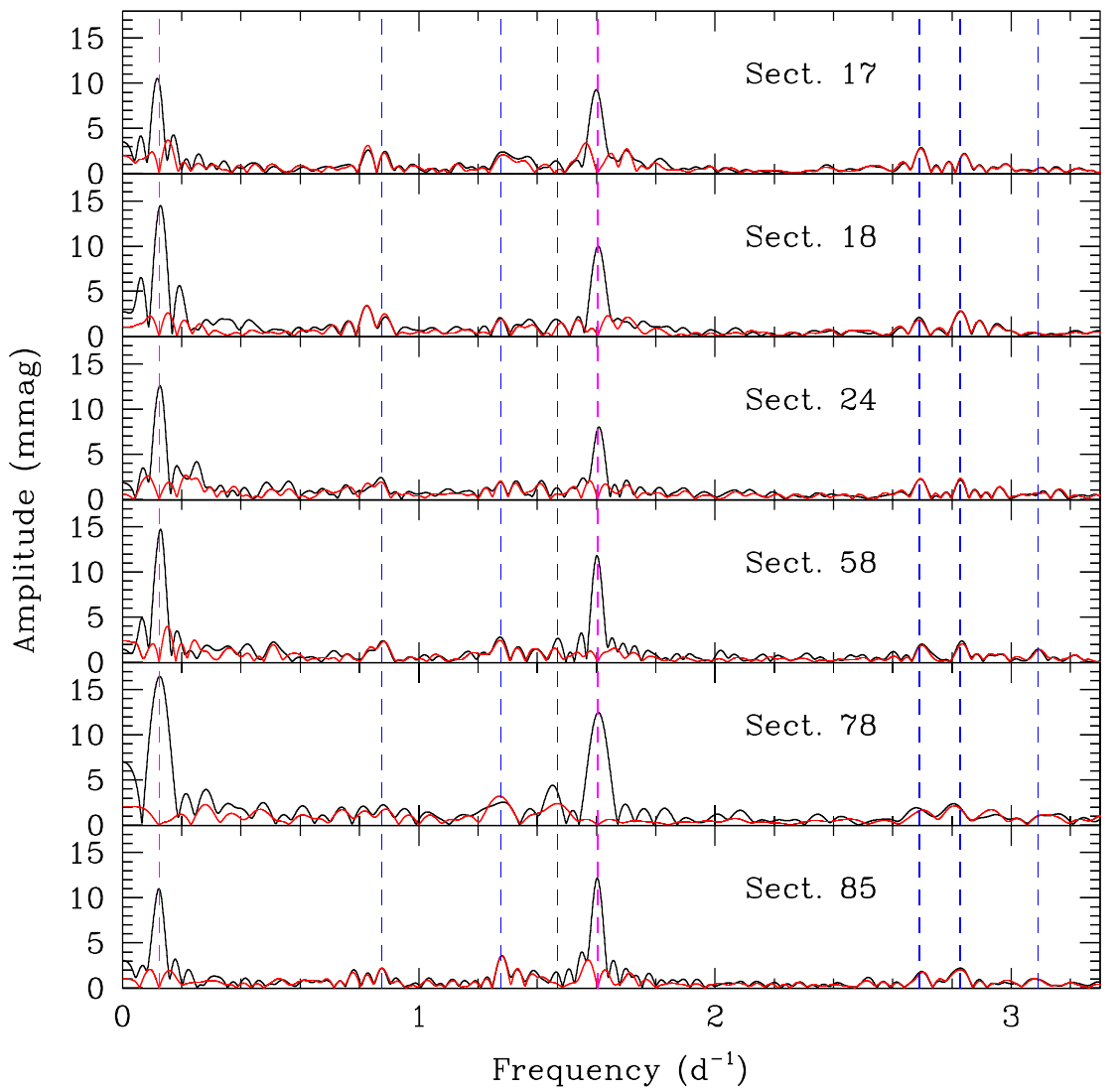}}
      \caption{Fourier periodogram of the {\it TESS} light curves of HD\,3191 as observed during Sectors 17, 18, 24, 58, 78 and 85 (black curves from top to bottom). The red curves show the periodograms after prewhitening the $\nu_1$ and $\nu_2$ frequencies. Dashed vertical lines yield the eight frequencies listed in Table\,\ref{tab_comb}: $\nu_1$ and $\nu_2$ are shown by magenta lines, whilst the other six frequencies are shown in blue.\label{TESSFour}}
    \end{center}
\end{figure}

Figure\,\ref{TESSlc} displays the {\it TESS} lightcurve during the six sectors.
One can clearly see two prominent modulations on timescales of about 0.6 and 8\,days as reported previously for Sectors 17, 18, and 24 by \citet{Mar21}.          
To further quantify these results, we applied the Fourier method of \citet{HMM}, amended by \citet{Gos01}, to compute the periodogram of these time series. This method explicity accounts for the oddities of the sampling of astronomical time series. Because of the different cadences of the time series (30\,min for Sector 17, 18, and 24, 200\,s for Sectors 58, 78, and 85), we first applied this method to the data from each individual sector separately. The results are displayed in Fig.\,\ref{TESSFour}. We computed the periodograms with frequency steps of $10^{-3}$\,d$^{-1}$ up to the theoretical Nyquist frequencies (25\,d$^{-1}$ for 30\,min cadence and 216\,d$^{-1}$ for 200\,s cadence). However, no significant power is found in the periodograms above about 3\,d$^{-1}$.

Two individual peaks, associated with the main modulations seen in the light curves, are consistently seen at all epochs. Their properties are summarised in Table\,\ref{tab_peaks}. At each epoch, the two strongest peaks are found at the same frequencies within the uncertainties. The peak at $\nu_1 = (0.125 \pm 0.003)$\,d$^{-1}$ corresponds to a period of almost exactly 8\,d. The second strongest peak has $\nu_2 = (1.604 \pm 0.003)$\,d$^{-1}$ which corresponds to a period of $0.623 \pm 0.002$\,d or $14.96 \pm 0.04$\,hr. The amplitudes of both peaks apparently change with time. 
The individual power spectra also hint at some additional peaks near $\nu_3 = (2.692 \pm 0.007)$\,d$^{-1}$ and $\nu_4 = (2.828 \pm 0.011)$\,d$^{-1}$. Although they are of much lower amplitude, they are seen in the periodograms of each individual sector and thus likely constitute persistent features. 

\begin{table*}[h!]
  \caption{Properties of the persistent peaks in the periodogram of the {\it TESS} light curves. The Fourier analysis was performed on the pca photometry, except for Sector 78 for which the median photometry was used.\label{tab_peaks}}
  \begin{center}
  \begin{tabular}{c c c c c c c c c c}
    \hline
    Sector & $\Delta \nu_{\rm nat}$ & $\nu_1$ & $A_1$ & $\nu_2$ & $A_2$ & $\nu_3$ & $A_3$ & $\nu_4$ & $A_4$\\
           & (d$^{-1}$) & (d$^{-1}$) & mmag & (d$^{-1}$) & mmag & (d$^{-1}$) & mmag & (d$^{-1}$) & mmag\\  
    \hline
    17 & 0.040 & 0.118 & 10.5 & 1.599 &  9.3 & 2.696 & 2.9 & 2.841 & 2.2\\
    18 & 0.042 & 0.128 & 14.5 & 1.606 & 10.0 & 2.689 & 2.1 & 2.830 & 2.8\\
    24 & 0.038 & 0.126 & 12.6 & 1.608 &  8.0 & 2.694 & 2.2 & 2.828 & 2.1\\
    58 & 0.036 & 0.128 & 14.7 & 1.602 & 11.8 & 2.698 & 2.0 & 2.833 & 2.3\\
    78 & 0.065 & 0.126 & 16.5 & 1.606 & 12.5 & 2.679 & 1.9 & 2.805 & 2.4\\
    85 & 0.039 & 0.123 & 11.0 & 1.603 & 12.2 & 2.698 & 1.8 & 2.828 & 2.2\\
    \hline
  \end{tabular}
  \end{center}
\end{table*}

\begin{figure}[h]
    \begin{center}
          \resizebox{8.5cm}{!}{\includegraphics{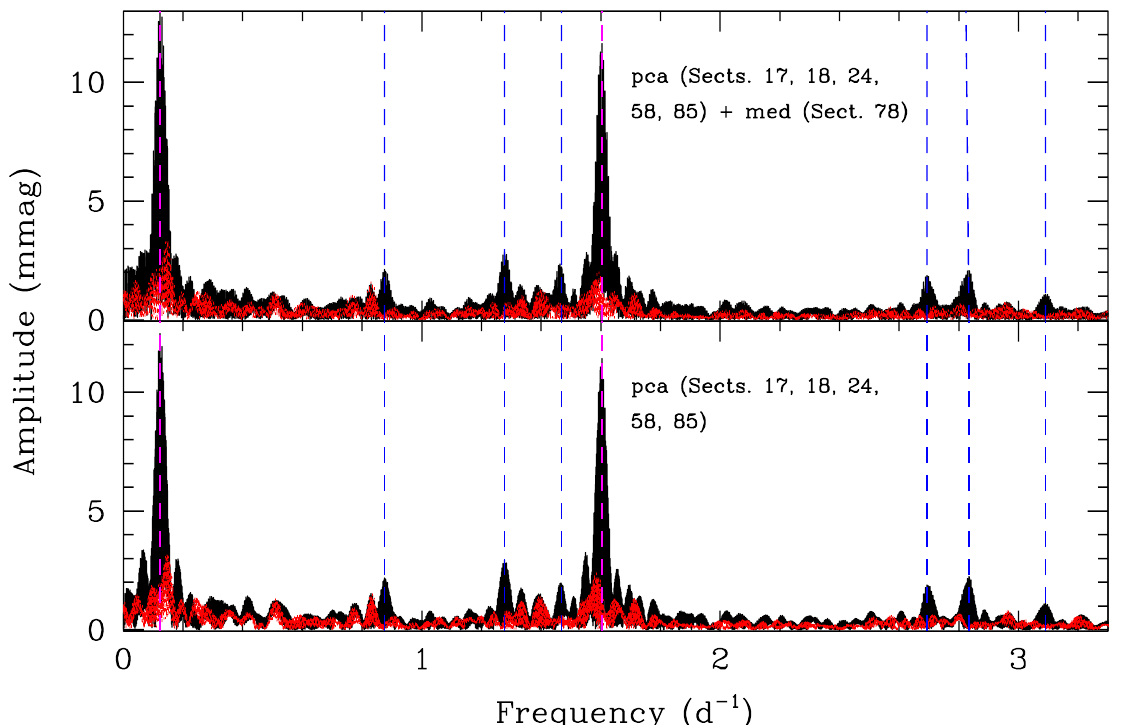}}
      \caption{Top panel: Fourier periodogram of the combined {\it TESS} data of all six sectors (black curve). The red curve yields the Fourier periodogram after prewhitening the eight frequencies listed in Table\,\ref{tab_comb} and indicated by the dashed vertical lines. Bottom panel: same for the combined light curves of those five sectors for which only pca photometry was used. \label{FourTESScomb}}
    \end{center}
\end{figure}

We also analysed the combined data from all sectors as well as from all sectors except Sector 78. The results are shown in Fig.\,\ref{FourTESScomb}. Due to the combination of data from different epochs, the power spectrum suffers from severe aliasing and the peaks consist of densely-packed narrow subpeaks. The frequencies of the highest subpeaks for $\nu_1$ and $\nu_2$ agree within $0.5\,\sigma$ with the mean values inferred above. Beside the main peaks, this analysis revealed several lower order peaks (see the lower half of Table\,\ref{tab_comb}). Some of them might be present in the power spectra of the individual sectors, but their visibility could be affected by red noise variations.
If $\nu_1$ reflects orbital motion in an eccentric binary, it could excite stellar oscillation modes via tidal excitation \citep[e.g.,][]{Kol23}. Some of the peaks could indeed be high order overtones of $\nu_1$, although given the low value of $\nu_1$, we must be cautious as these near integer ratios could be fortuitous. For instance, the $\nu_2$ frequency itself is consistent with 13 times $\nu_1$, but this relation only holds for the frequencies listed in Table\,\ref{tab_comb} determined from the Fourier analyses combining at least five {\it TESS} sectors. It does not work for the (less accurate) mean frequencies determined from the individual sectors. Likewise $\nu_4$ is consistent with 23 times $\nu_1$, and the $\nu_8$ frequency matches $25\,\nu_1$. Conversely, there exists no simple integer ratio with $\nu_1$ for $\nu_3$ (close to $21.8\,\nu_1$), $\nu_5$ (close to $7.1\,\nu_1$), $\nu_6$ (close to $10.3\,\nu_1$), and $\nu_7$ (close to $11.9\,\nu_1$). This is also valid for any linear combination of $\nu_1$ and $\nu_2$.

\begin{table}
  \caption{Properties of the strongest peaks in the periodogram of the combined {\it TESS} data.\label{tab_comb}}
  \begin{center}
  \begin{tabular}{c c c c c}
    \hline
    Designation & \multicolumn{2}{c}{pca (5 sect.)} & \multicolumn{2}{c}{pca (5 sect.)} \\
                & \multicolumn{2}{c}{+ med (Sect.\ 78)} & \\
        & $\nu$ & $A$ & $\nu$ & $A$\\
           & (d$^{-1}$) & mmag & (d$^{-1}$) & mmag \\  
    \hline
    $\nu_1$ & 0.1235 & 13.1 & 0.1235 & 12.5 \\
    $\nu_2$ & 1.6038 & 11.6 & 1.6038 & 11.4 \\
    $\nu_3$ & 2.6933 &  1.9 & 2.6933 &  1.9 \\
    $\nu_4$ & 2.8349 &  2.1 & 2.8349 &  2.2 \\    
    $\nu_5$ & 0.8750 &  2.1 & 0.8750 &  2.2 \\
    $\nu_6$ & 1.2780 &  2.9 & 1.2780 &  2.9 \\
    $\nu_7$ & 1.4691 &  2.2 & 1.4664 &  2.0 \\
    $\nu_8$ & 3.0910 &  1.1 & 3.0911 &  1.1 \\
    \hline
  \end{tabular}
  \end{center}
\end{table}

\begin{figure}[h]
    \begin{center}
          \resizebox{8.5cm}{!}{\includegraphics{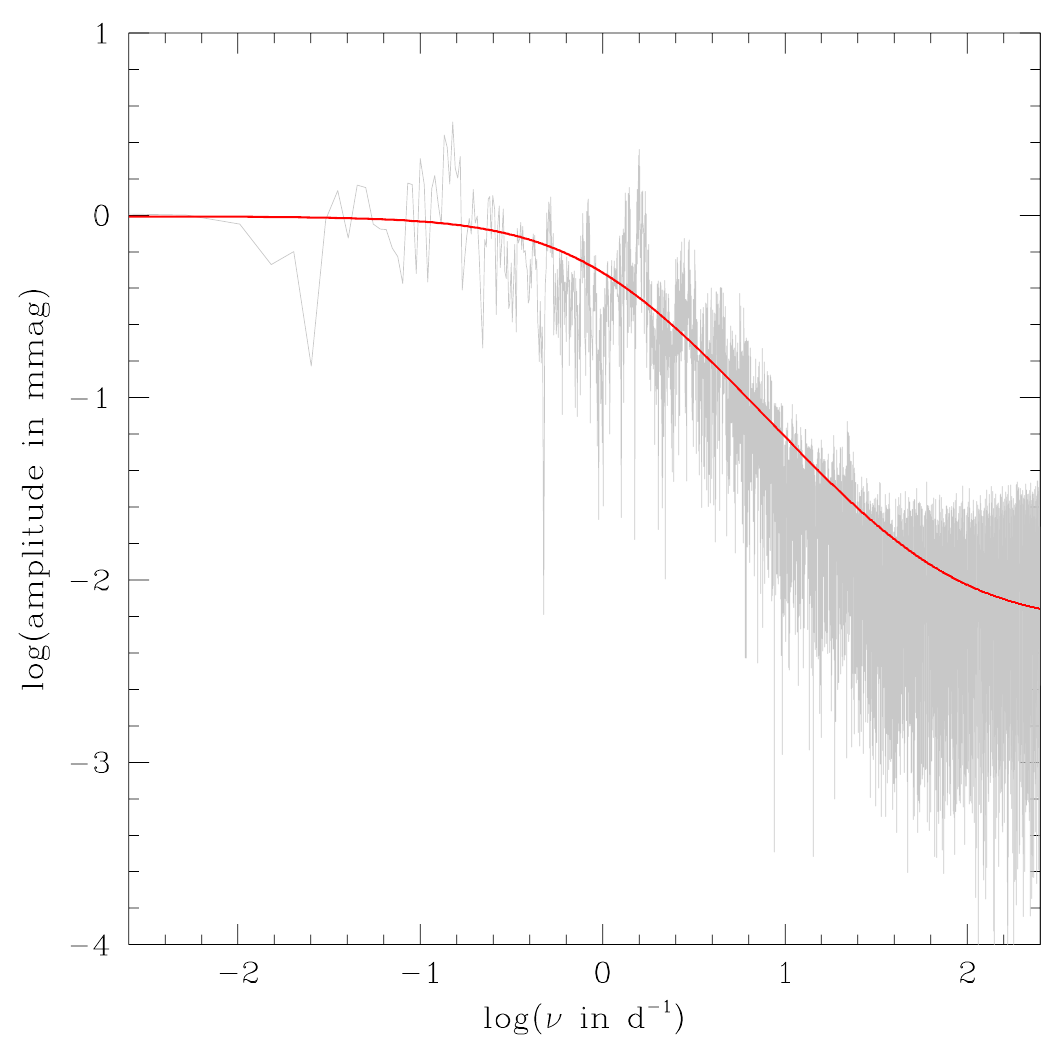}}
      \caption{Fourier periodogram of the residuals of the {\it TESS} light curves of Sectors 58 and 85 after prewhitening the variations with the eight frequencies quoted in Table\,\ref{tab_comb}. The red curve yields the best-fit red noise model adopting the formalism of \citet{Sta02} and the fitting procedure of \citet{Naz21}. \label{rednoise}}
    \end{center}
\end{figure}

We prewhitened the photometric time series by subtracting expressions of the kind
\begin{equation}
  \sum_{n=1}^{N} A_n\,\sin{(2\,\pi\,\nu_n\,t + \phi_n^0)},
\end{equation}
where $A_n$ and $\phi_n^0$ are the amplitude and phase constant of frequency $\nu_n$ adjusted by a least-square technique to the observed data. The results of the prewhitening procedure with the $\nu_1$ and $\nu_2$ frequencies (from Table\,\ref{tab_comb}) applied to the data from the individual sectors is displayed by the red curves in Fig.\,\ref{TESSFour}. The red periodogram in Fig.\,\ref{FourTESScomb} shows the results after prewhitening all eight frequencies quoted in Table\,\ref{tab_comb}. Residuals at the few mmag level are seen near $\nu_1$ and $\nu_2$. These residuals likely stem from the fact that the amplitudes of individual sectors slightly differ from the values in Table\,\ref{tab_comb}. The prewhitened light curve (after removing the eight frequencies from Table\,\ref{tab_comb}) still displays residual variations. Figure\,\ref{rednoise} illustrates the periodogram of the combined data from Sectors 58 and 85, which have the highest Nyquist frequency among those with pca photometry, in a log-log scale. This unveils the presence of a red noise component in the periodogram as is commonly found in high-precision photometry of all kinds of massive stars \citep[e.g.,][]{Blo11,Bow20,Naz21,Naz24,She24}.

\subsection{Stellar parameters \label{SpT}}
In the compilation of spectral types and photometric data of \citet{Ree03}, HD\,3191 is assigned a B1\,IV:nn spectral type, with the colon and nn tags indicating respectively an uncertain luminosity class and very broad absorption lines. Analysing the mean TIGRE/HEROS spectrum of HD\,3191 confirms the B1\,IV:nn  spectral classification and yields a projected rotational velocity $v\,\sin{i} = (256 \pm 5)$\,km\,s$^{-1}$ (see \ref{appSpT}).

At this stage, we emphasise that none of our spectra displays an He\,{\sc ii} $\lambda$\,4686 emission, unlike what \citet{Mar21} reported. If related to an accretion structure around a putative companion such an emission could be transient or appear only at certain orbital phases (especially for an eccentric binary). Regarding the last point, we stress that our spectra sample the proposed orbital period rather well. Therefore, unless the emission would only be visible over a very short phase interval, it seems very unlikely that our observations could have missed it. The spectrum presented by \citet{Mar21} was of very low resolution (resolving power $\sim 1100$) and limited signal-to-noise ratio. In view of our high-resolution data, it seems that the weak feature they reported was not of astrophysical origin.

\citet{Pan21} inferred a distance of $1194^{+50}_{-60}$\,pc using {\it Gaia}-DR2 data, whilst \citet{Bai21} obtained $1181^{+19}_{-18}$\,pc using {\it Gaia}-eDR3 astrometry. In \ref{app2}, we show that the star is surrounded by a dusty IR nebula, which is responsible for part of the star's reddening. Together with $m_V = 8.58 \pm 0.01$ \citep{Ree03} and $A_V = 2.16\pm 0.09$ (see \ref{app2}), the distance $(1187 \pm 55)$\,pc (i.e.\ the mean of the above values) leads to an absolute $V$-band magnitude of $-3.95 \pm 0.10$. \citet{Hum84} adopted absolute magnitudes of $-3.2$, $-3.8$, and $-4.3$ respectively for B1\,V, B1\,IV and B1\,III stars. Our derived magnitude agrees well with the value for a subgiant as quoted by these authors\footnote{\citet{Weg06} derived absolute magnitudes of OB stars using {\it Hipparcos} parallaxes and inferred $M_V$ values of $-2.95 \pm 1.6$, $-3.28 \pm 1.1$ and $-4.10 \pm 2.2$ for B1-1.5 stars of respectively main-sequence, subgiant and giant luminosity class. Given the rather large uncertainties on the calibrations of \citet{Weg06}, our value of the absolute magnitude is consistent with any of the luminosity classes V, IV or III.}. 

\citet{Nie13} derived bolometric corrections of $-2.65$ and $-2.67$ respectively for B1\,V and B1\,IV stars, whilst \citet{Pec13} give $-2.61$ for B1\,V stars. Assuming an error of 0.05 on the bolometric correction (adopted as $-2.64$, i.e., the mean of the above values), we obtain an estimate of the bolometric luminosity of HD\,3191 of $L_{\rm bol} = (1.31 \pm 0.13) \times 10^{38}$\,erg\,s$^{-1}$ or $(34040 \pm 3450)$\,L$_{\odot}$. This bolometric luminosity is $\sim 1.8$ times larger than the value used by \citet{Mar21}. The reason is that these authors did not compute $M_V$, but instead adopted the value from \citet{Weg06} for B1-1.5\,IV stars. Assuming an uncertainty on $T_{\rm eff}$ of 1000\,K, our estimate of the bolometric luminosity yields a stellar radius of $(8.46 \pm 0.76)$\,R$_{\odot}$.

  Adopting $T_{\rm eff}$ and $L_{\rm bol}$ inferred above, we used the {\tt BONNSAI} tool \citep{Sch17} to estimate the mass and age of HD\,3191. The results are an initial mass of $14.8 \pm 0.6$\,M$_{\odot}$ and an age of $8.8 \pm 0.7$\,Myr, based on rotating single-star evolutionary tracks from \citet{Bro11} at metallicity $Z = 0.009$. In the following, we thus assume a stellar mass of 14.8\,M$_{\odot}$ as obtained from {\tt BONNSAI}, but we stress that all our conclusions remain fully valid for slightly lower masses (e.g., adopting 12\,M$_{\odot}$).

\begin{figure}[h]
    \begin{center}
          \resizebox{8.5cm}{!}{\includegraphics{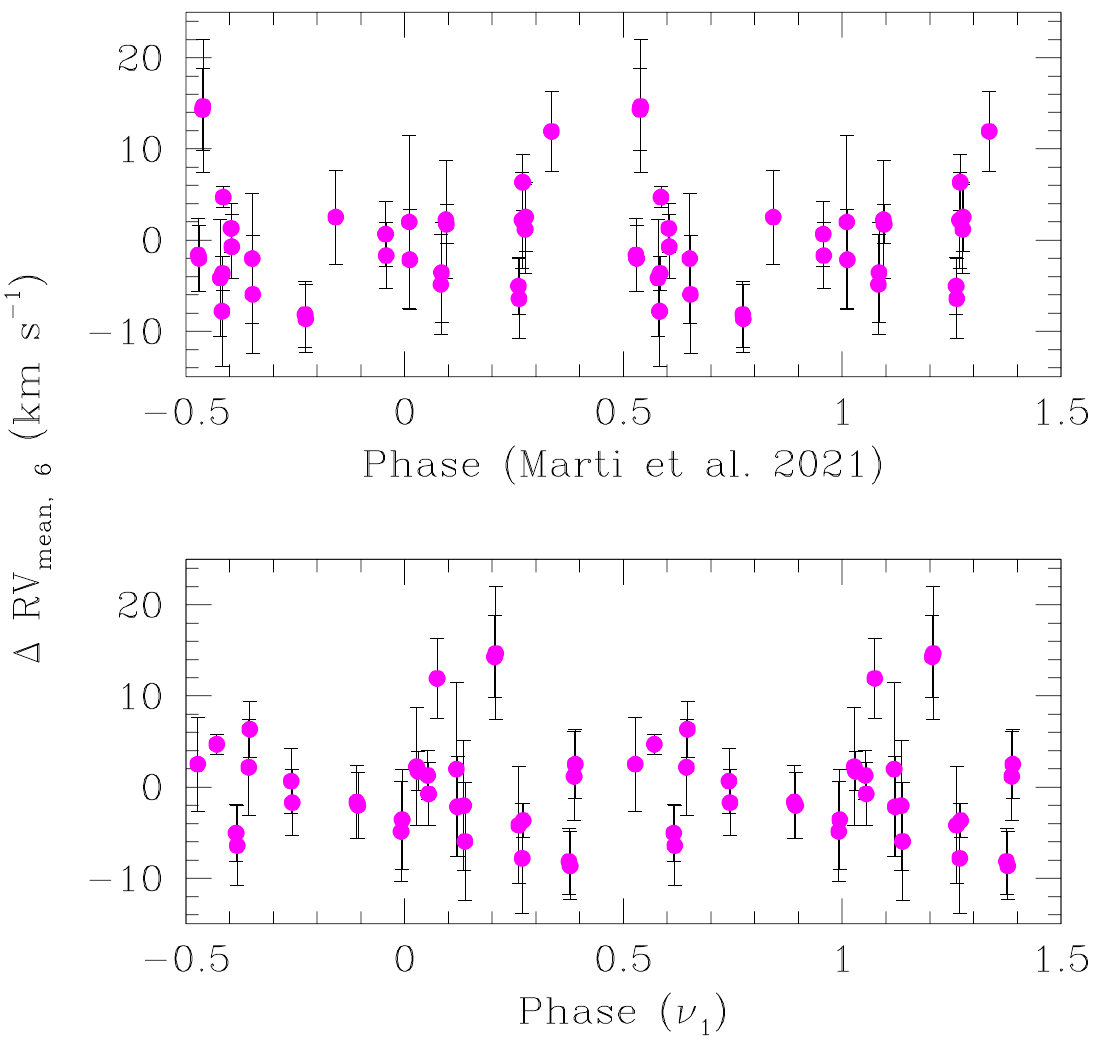}}
      \caption{RV$_{\rm mean, 6}$ as a function of phase according to the ephemerides of \citet{Mar21} (top) or our $\nu_1$ frequency (bottom).\label{FigRVmean}}
    \end{center}
\end{figure}
\subsection{Radial velocities}
           Owing to the rapid rotation, the spectral lines of HD\,3191 are heavily broadened. This situation directly impacts the measurement of the star's RVs. We performed such measurements using different techniques outlined in \ref{app1}. The best estimates were obtained for RV$_{\rm mean,6}$, that is the mean RV of the six strongest spectral lines (H$\gamma$, He\,{\sc i} $\lambda$\,4471, H$\beta$, He\,{\sc i} $\lambda$\,5876, H$\alpha$, and He\,{\sc i} $\lambda$\,6678).  

           Folding the values of RV$_{\rm mean, 6}$ with the ephemerides of \citet{Mar21} or with our $\nu_1$ frequency, no obvious trend attributable to an orbital motion appears. Instead, the RVs from neighbouring phases scatter randomly (see Fig.\,\ref{FigRVmean}). This result contrasts with the conclusion of \citet{Mar23} who reported hints of a 16.09\,d period in their RV data. These authors collected 29 spectra of HD\,3191 in the 6300 -- 6730\,\AA\ wavelength range at a resolving power of 12000. Their RVs were measured by cross-correlation with a synthetic spectrum. Applying a Phase Dispersion Minimisation method \citep[PDM,][]{Ste78}, they reported a most likely period in the 15.9 -- 16.5\,d interval. Assuming a circular orbit, their best-fit RV amplitude was $K_{\rm Marti} = (5.2 \pm 1.4)$\,km\,s$^{-1}$.

\begin{figure*}[h]
    \begin{center}
      \resizebox{5.6cm}{!}{\includegraphics{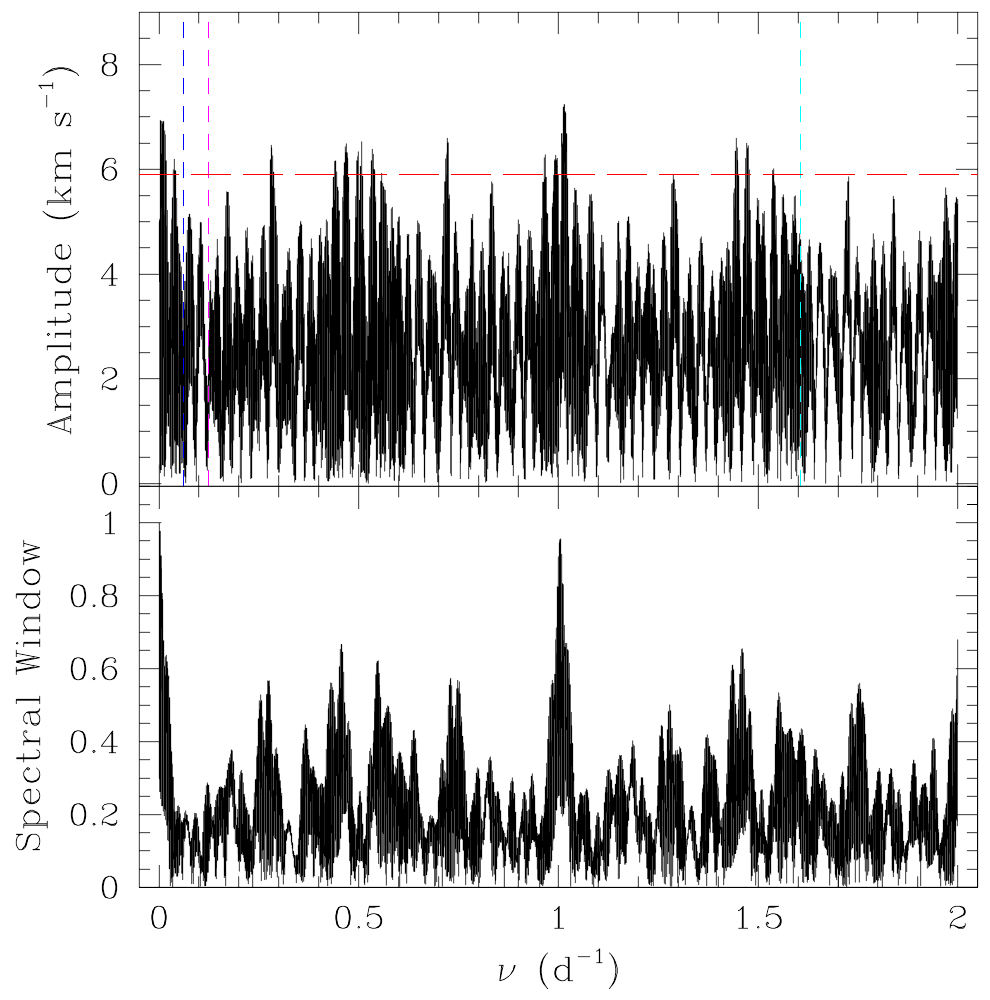}}
      \resizebox{5.6cm}{!}{\includegraphics{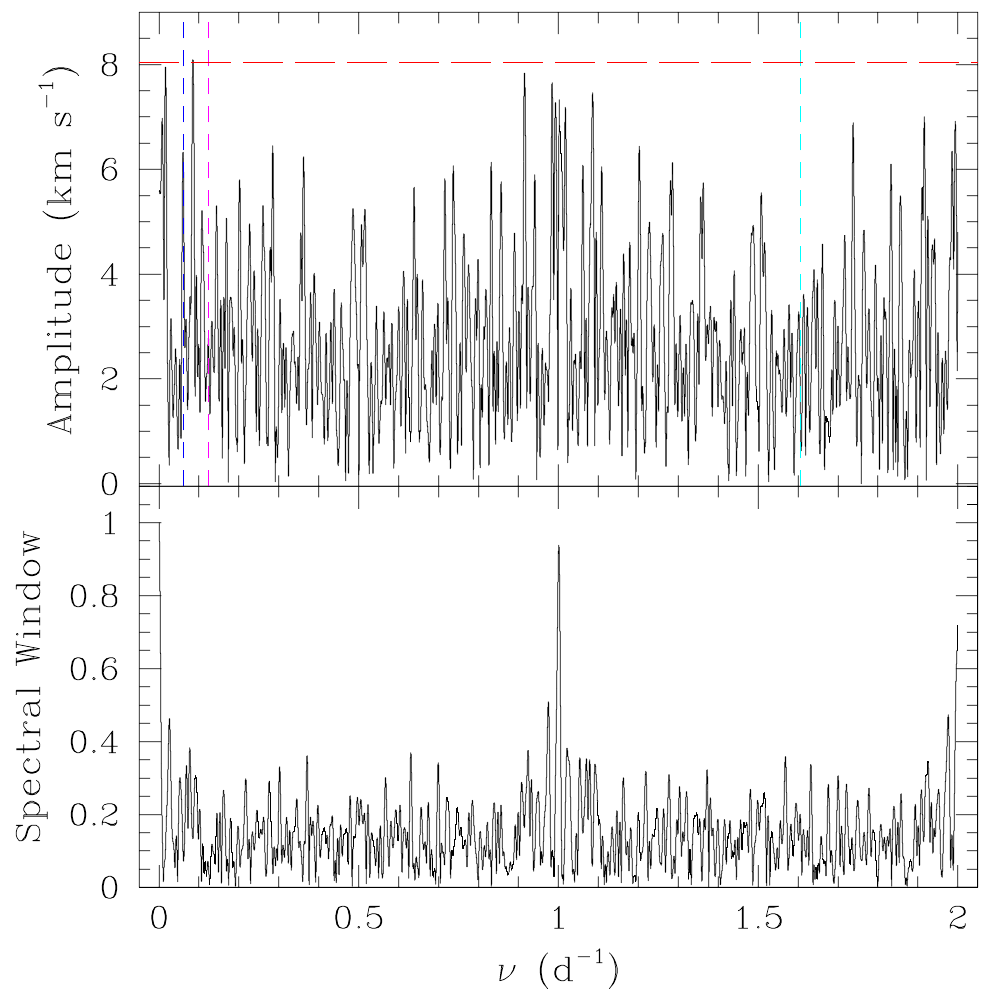}}
      \resizebox{5.6cm}{!}{\includegraphics{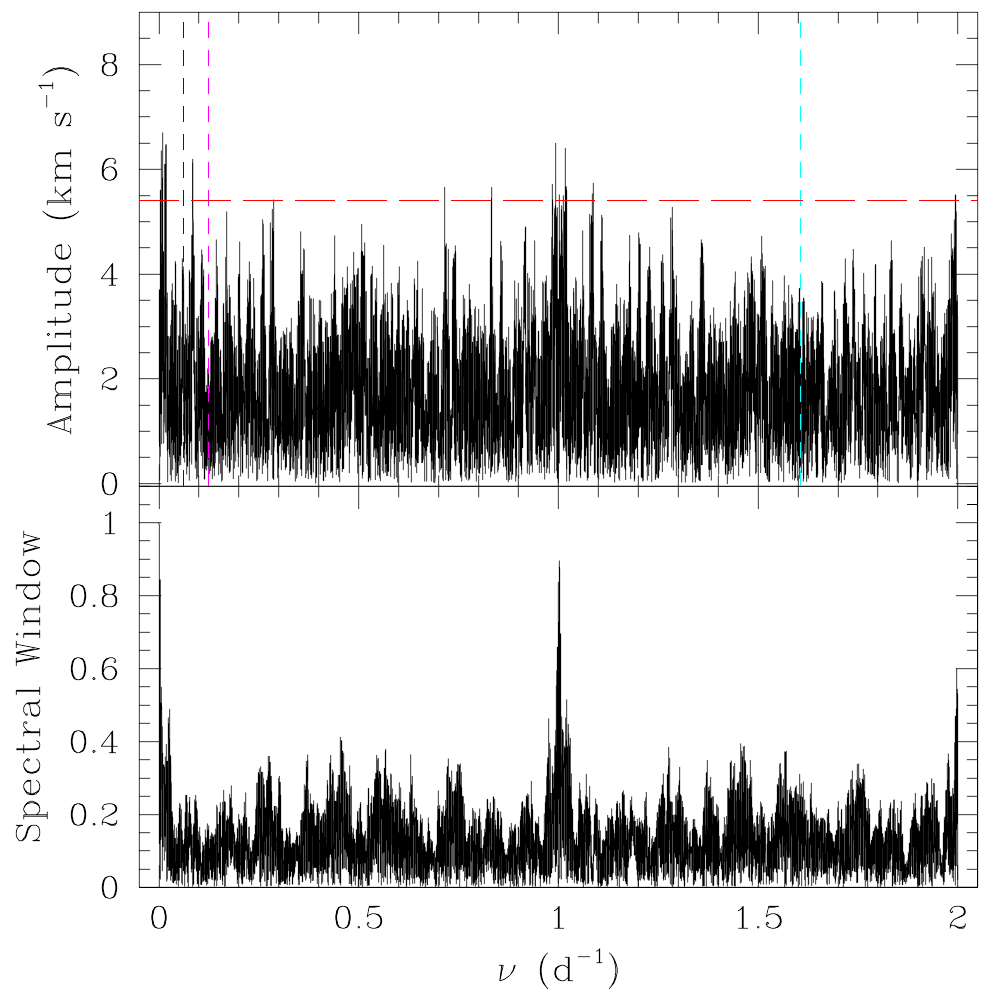}}      
      \caption{From left to right: Fourier periodogram and spectral window of the RV$_{\rm mean, 6}$ data, the RVs from \citet{Mar23}, and the combined RV dataset. The blue, magenta and cyan vertical dashed lines indicate respectively $\nu_1/2$, $\nu_1$ and $\nu_2$. The red horizontal dashed line yields the 1\% significance level (see text).\label{FigFourRVs}}
    \end{center}
\end{figure*}

Figure\,\ref{FigFourRVs} illustrates the Fourier periodograms, computed with the method of \citet{HMM} and \citet{Gos01}, of our 30 RV$_{\rm mean, 6}$ data, of the 29 RVs taken from \citet{Mar23}, and of our new data combined with those of \citet{Mar23} after subtracting the mean values from each dataset. To assess the significance level of the features in the power spectrum, we applied a bootstrapping method where the observation dates were kept fixed but the RVs were randomly redistributed among these dates. Each reshuffled time series was injected into the Fourier method and the amplitude of the strongest peak was recorded. For each of the three time series (RV$_{\rm mean, 6}$, RV$_{\rm Marti}$, RV$_{\rm combined}$), we performed 1000 reshuffling trials. The histogram of the highest peaks in the periodograms was used to determine the amplitude corresponding to a 1\% significance level, that is 1\% of the periodograms of the randomly reshuffled time series produce a peak above this threshold. For each time series, this threshold is illustrated by the red horizontal dashed line in Fig.\,\ref{FigFourRVs}.

The Fourier periodograms of the RV time series essentially reflect the convolution of some low amplitude variations at low frequency with the spectral window. Whilst several peaks exceed the 1\% significance level, this is not the case for either of $\nu_1/2$, $\nu_1$, or $\nu_2$. In the periodogram of the RVs of \citet{Mar23}, the strongest peak is found at a frequency of $0.08498$\,d$^{-1}$, corresponding to a period of 11.77\,d and a RV semi-amplitude of 8.1\,km\,s$^{-1}$. Moreover, we note that the value of $K_{\rm Marti} = (5.2 \pm 1.4)$\,km\,s$^{-1}$ is well below the 1\% significance level for their dataset. Generally speaking, no significant signature of the strongest photometric signals is found among the RVs. 

\begin{figure}[h]
    \begin{center}
          \resizebox{8.5cm}{!}{\includegraphics{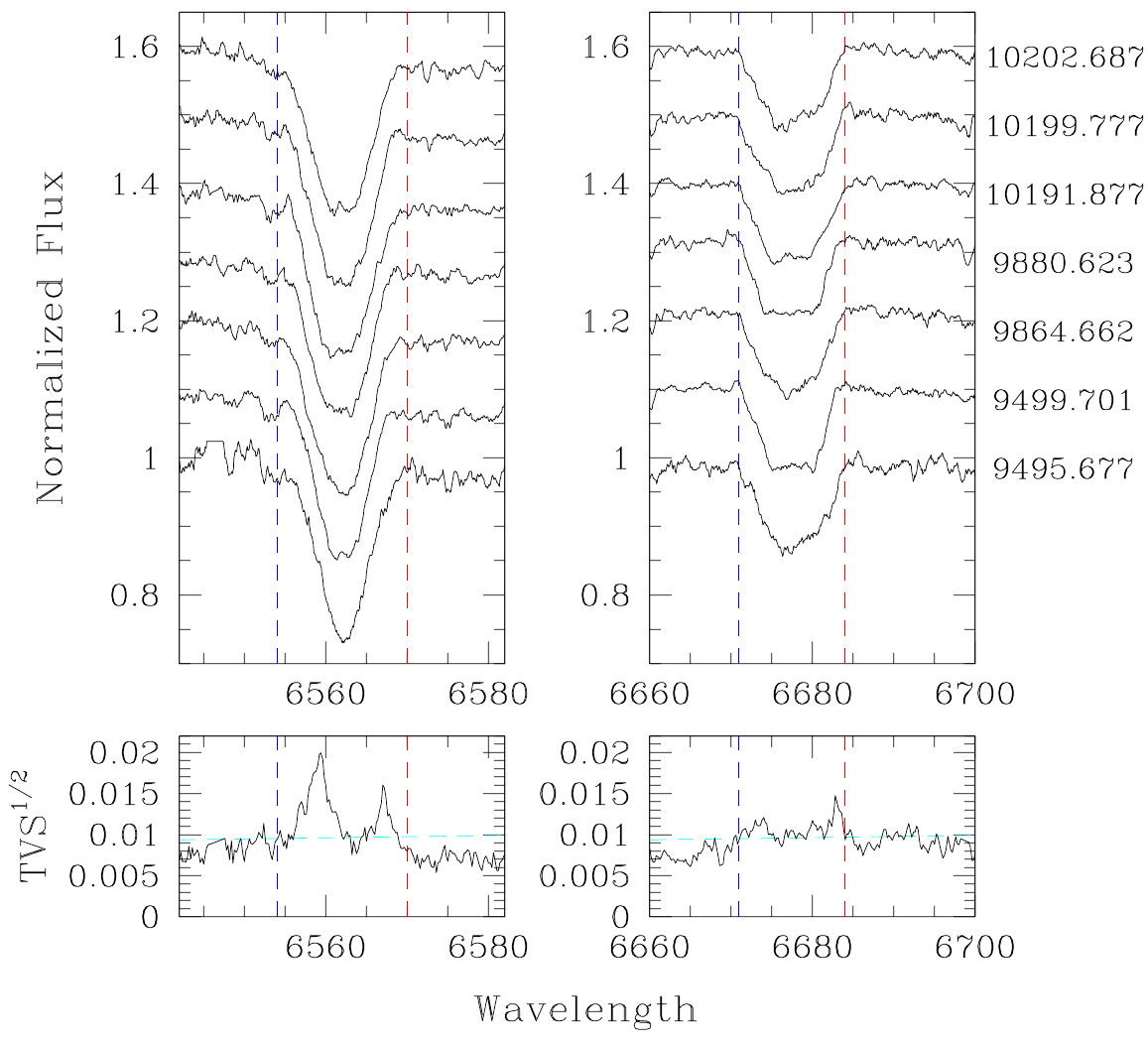}}
          \caption{Top: Montage of some representative TIGRE/HEROS observations of the H$\alpha$ (left) and He\,{\sc i} $\lambda$~6678 (right) lines in the spectrum of HD\,3191. The dates (in the format HJD - 2450000) are indicated on the right of the plot. Whilst no obvious overall shifts in wavelength are seen, the lines clearly underwent profile variations which most probably stem from NRPs. Bottom: temporal variance spectrum in the same wavelength regions. The cyan dashed line corresponds to the 99\% significance level.\label{Figlpv}}
    \end{center}
\end{figure}

Taking the combined dataset, the $1\sigma$ dispersion of the RVs amounts to 7.2\,km\,s$^{-1}$. The likely origin of the RV variations becomes clearer by examining the line profiles. Indeed, Fig.\,\ref{Figlpv} reveals line profile variations (lpvs) affecting the rotationally broadened profile of the H$\alpha$ and He\,{\sc i} $\lambda$\,6678 lines. Though our TIGRE  spectroscopic time series is not designed to probe the timescale of the $\nu_2$ photometric signal of HD\,3191, the most likely explanation of these lpvs and of the RV dispersion is that they stem from non-radial pulsations (NRPs) of the star.

The bottom panel of Fig.\,\ref{Figlpv} illustrates the temporal variance spectrum \citep{Ful96} of the TIGRE spectra computed in the region around the H$\alpha$ and He\,{\sc i} $\lambda$\,6678 lines. One can clearly see a strong double-peaked variability feature in the H$\alpha$ line, and to a lesser extent also in the He\,{\sc i} $\lambda$\,6678 line. This feature could either reflect a genuine orbital motion, or lpvs strongly affecting the wings of the lines as expected for NRPs with relatively high values of the ratio of the horizontal-to-radial velocity amplitude \citep{Kam88}. The latter scenario seems supported by comparing the line profiles in the top panel of Fig.\,\ref{Figlpv} with Fig.\,5 of \citet{Kam88}.   

The absence of significant RV variations prevents us from computing an orbital solution. In principle, we could nevertheless attempt to establish such a solution via the light travel time effect. Indeed, the pulsations of a star moving around the centre of mass of a binary system undergo a light arrival time delay effect that is modulated by the orbital motion as
\begin{equation}
  \tau = -\frac{a\,\sin{i}}{c}\,(1 - e^2)\,\frac{\sin{(\phi(t)+\omega})}{1 + e\,\cos{\phi(t)}}.
\end{equation}
In this relation $a$ is the semi-major axis of the pulsating star's orbit around the centre of mass, $i$ is the orbital inclination, $e$ the eccentricity, $\phi(t)$ the true anomaly at time $t$ and $\omega$ the argument of periastron. The time delay can be used to establish the properties of the orbit \citep[e.g.,][]{Mur15,Lam21}. The projected semi-major axis, $a\,\sin{i}$, is directly connected to the semi-amplitude of the radial velocity curve, $K$, via
\begin{equation}
  a\,\sin{i} = \frac{K\,P_{\rm orb}\,\sqrt{1 - e^2}}{2\,\pi}.
\end{equation}
Taking a value of $K \leq 21.6$\,km\,s$^{-1}$, corresponding to the $3\,\sigma$ upper limit inferred above, yields $\frac{a\,\sin{i}}{c} \leq 16$\,s. This value is significantly smaller than the time resolution of the {\it TESS} data.

We attempted nonetheless to check for delays in the times of the photometric minima associated with the $\nu_2$ frequency. To do so, we first prewhitened the data from Sectors 58 and 85 for the other seven frequencies from Table\,\ref{tab_comb}, that is all frequencies except $\nu_2$. In this way, the residual photometric modulations should be dominated by the $\nu_2$ frequency. We then adjusted the times of minima by fitting a parabola to the data points falling in the $\nu_2$ phase interval 0.8 to 1.2.

\begin{figure}[h]
    \begin{center}
          \resizebox{8.5cm}{!}{\includegraphics{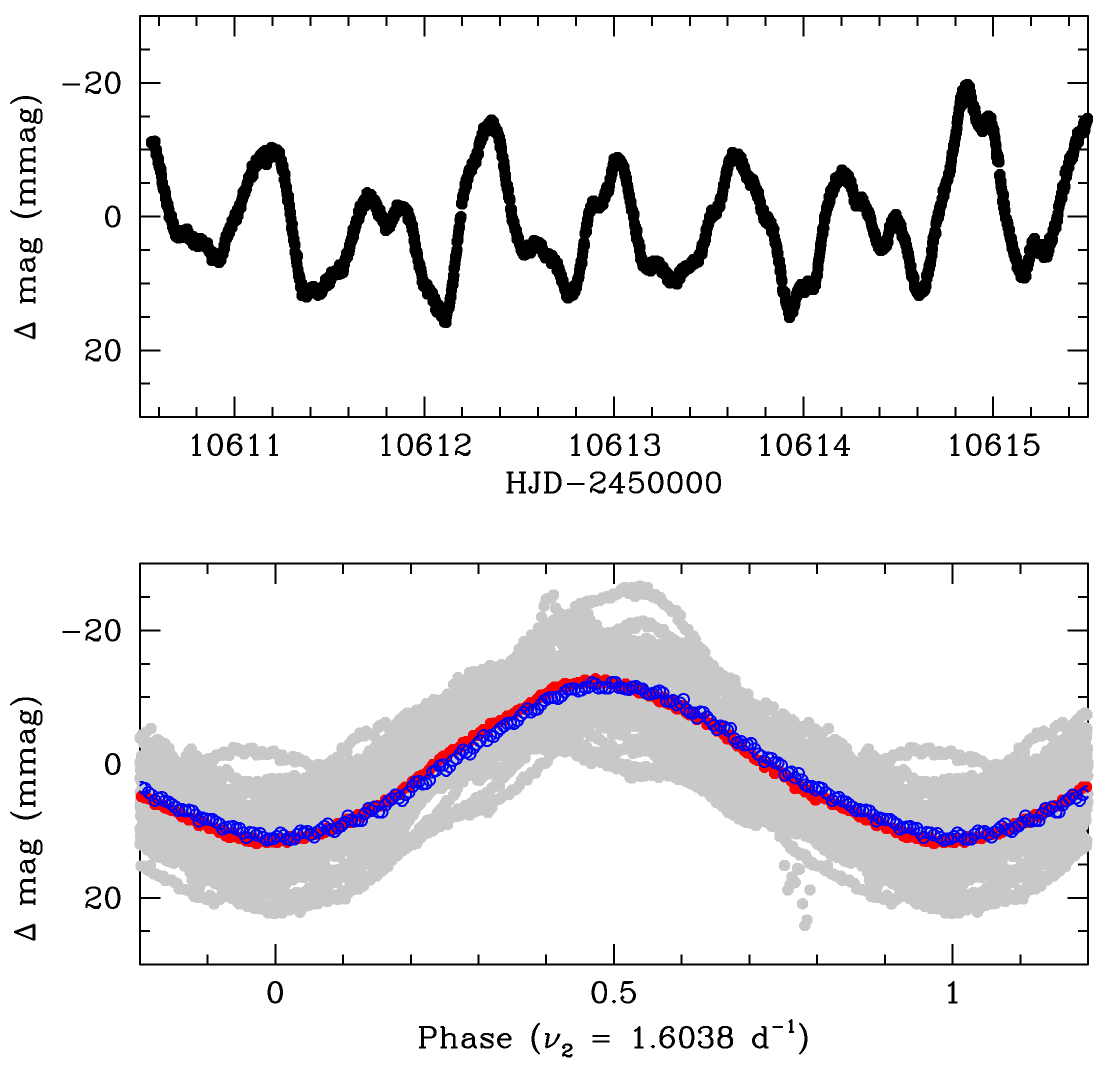}}
          \caption{Properties of the variations at the $\nu_2$ frequency. The top panel shows a zoom on five days of Sector 85 showing the photometric residuals after prewhitening seven frequencies other than $\nu_2$. Despite removing the most significant modes other than $\nu_2$, the signal displays a complex morphology. The bottom panel displays the same residual photometry of Sector 85 folded in phase with $\nu_2$ (grey symbols) and taking $t_0$ equal to HJD\,$2\,460\,610.857612$. The red symbols illustrate the mean variations at the $\nu_2$ frequency during Sector 85, obtained by averaging the data in 200 equal phase bins. The open blue symbols yield the same mean variations computed using the residuals of all {\it TESS} sectors; they agree well with the results of Sector 85.\label{pulsations}}
    \end{center}
\end{figure}

If the star describes an orbital motion with a period of 16.09\,days, then the delays of the times of minima with respect to linear ephemerides should display a systematic modulation \citep[e.g.][]{Mur15}. This is not the case here. Moreover, the measured delays span a rather wide range, between $-0.05$\,d and $0.05$\,d for Sect.\,85 and even more for Sect.\,58, which is clearly inconsistent with our RV data. A close inspection of the shapes of the residual light curve (Fig.\,\ref{pulsations}) reveals that, despite the oscillation having on average a sinusoidal shape, the individual cycles display complex and changing morphologies. This jeopardises any attempts to detect the subtle delays that could be due to light travel time effects. 

\begin{table*}
  \caption{Results of our analysis of the January 2016 EPIC-pn spectrum of HD\,3191. The observed and absorption-corrected fluxes respectively in the sixth and seventh columns are evaluated over the 0.5 - 10\,keV energy range. The norm of the powerlaw model corresponds to the photon flux per 1\,keV energy range at an energy of 1\,keV. The norm of the {\tt apec} model corresponds to $\frac{\int n_e\,n_{\rm H}\,dV}{4\,\pi\,d^2}$, where $d$ is the distance to the source, $n_e$ and $n_{\rm H}$ are the electron and proton density of the emitting plasma.\label{fitX}}
  \begin{center}
  \begin{tabular}{c c c c c c c}
      \hline
      \multicolumn{7}{c}{{\tt phabs*power}}\\
      $N_{\rm H}$ & $\Gamma$ & norm & $\chi^2_{\nu}$ & d.o.f. & $f_X^{\rm obs}$ & $f_X^{\rm unabs}$ \\
      $(10^{22}$\,cm$^{-2})$ & & (photons\,keV$^{-1}$\,cm$^{-2}$\,s$^{-1}$) & & & ($10^{-14}$\,erg\,cm$^{-2}$\,s$^{-1}$) & ($10^{-14}$\,erg\,cm$^{-2}$\,s$^{-1}$) \\
      \hline
      \vspace*{-3mm}\\
      $0.07^{+0.30}_{-0.07}$ & $0.94^{+0.68}_{-0.46}$ & $(5.7^{+3.3}_{-2.0}) \times 10^{-6}$ & 0.57 & 3 & $9.2^{+1.3}_{-2.1}$ & $9.4^{+1.9}_{-1.9}$ \\
      \vspace*{-3mm}\\
      $0.42$ (fixed) & $1.47^{+0.59}_{-0.52}$ & $(11.6^{+3.7}_{-3.2}) \times 10^{-6}$ & 1.25 & 4 & $7.7^{+1.8}_{-1.8}$ & $9.4^{+1.6}_{-1.2}$ \\
      \vspace*{-3mm}\\
      \hline
      \multicolumn{7}{c}{{\tt phabs*apec}}\\
        $N_{\rm H}$ & $kT$ & norm & $\chi^2_{\nu}$ & d.o.f. & $f_X^{\rm obs}$ & $f_X^{\rm unabs}$ \\
      $(10^{22}$\,cm$^{-2})$ & (keV) & ($10^{-14}$\,cm$^{-5}$) & & & ($10^{-14}$\,erg\,cm$^{-2}$\,s$^{-1}$) & ($10^{-14}$\,erg\,cm$^{-2}$\,s$^{-1}$) \\
      \hline
      \vspace*{-3mm}\\
      $0.15^{+0.26}_{-0.11}$ & $\geq 6.8$ & $(4.9^{+0.9}_{-1.4}) \times 10^{-5}$ & 0.84 & 3 & $7.9^{+1.0}_{-1.2}$ & $8.3^{+1.6}_{-1.5}$ \\
       \vspace*{-3mm}\\     
       $0.42$ (fixed) & $\geq 4.4$ & $(4.6^{+2.2}_{-0.2}) \times 10^{-5}$ & 1.33 & 4 & $7.7^{+1.9}_{-1.7}$ & $9.3^{+1.9}_{-1.5}$ \\
      \vspace*{-3mm}\\       
        \hline
  \end{tabular}
  \end{center}
\end{table*}

\subsection{X-ray properties}
We used the {\tt xspec} version 12.9.0i software \citep{Arn96} to analyse the January 2016 EPIC-pn spectrum of HD\,3191. Together with the relation between interstellar hydrogen column density and colour excess from \citet{Gud12}, the above inferred $E(B-V)$ colour excess yields an estimated hydrogen column density of $N_{\rm H} = 4.2 \times 10^{21}$\,cm$^{-2}$. We tested models of absorbed APEC \citep{Smi01} optically thin thermal plasma models or absorbed power law models to adjust the X-ray spectral energy distribution. We either requested the column density to be $4.2 \times 10^{21}$\,cm$^{-2}$, or let this parameter vary freely in the fitting procedure. 

The results are summarised in Table\,\ref{fitX}. In general, the power law models yield somewhat better quality adjustments than the APEC models. The APEC models result in poorly constrained plasma temperatures that are unusually high for X-ray spectra of OB stars \citep{Naz09,Naz18}. Models with the column density set to the estimated value resulted in poorer fitting qualities, especially at energies below 1\,keV where the model systematically underpredicted the observed spectrum. The discrepancy between the best-fit $N_{\rm H}$ and the value estimated from $E(B - V)$ most probably results from the dusty nebula towards HD\,3191 detected in \ref{app2}. This nebula significantly contributes to the reddening, but could have a lower gas content than the average interstellar medium.  

The most robust result from the X-ray analysis concerns the absorption-corrected flux, which is found to be $(9.1 \pm 0.8) \times 10^{-14}$\,erg\,cm$^{-2}$\,s$^{-1}$. This flux results in an X-ray luminosity of $L_{\rm X} = (1.54 \pm 0.20) \times 10^{31}$\,erg\,s$^{-1}$ and $\log{\frac{L_{\rm X}}{L_{\rm bol}}} = -6.93 \pm 0.07$. Both values would be extremely low for an HMXB, but are fully consistent with the range found for B-type stars observed with {\it XMM-Newton} \citep[$\log{\frac{L_{\rm X}}{L_{\rm bol}}} = -6.81 \pm 0.74$ for good quality data,][]{Naz09}. Likewise, {\it Chandra} observations of B-type stars in the Carina complex yielded a mean $L_{\rm X}$ of $1.4 \times 10^{31}$\,erg\,s$^{-1}$ over the 0.5 - 10\,keV energy band, and $kT$ values often exceeding 1.5\,keV \citep{Naz11}. A very similar picture (average $kT$ of 2.4\,keV and $L_{\rm X}$ near $10^{31}$\,erg\,s$^{-1}$) was found for B-type stars in the {\it Chandra} survey of the Cygnus OB2 association \citep{Rau15}. The X-ray properties of these B-type stars are consistent with the X-ray emission coming from a low-mass pre-main sequence star. We thus conclude that the existing X-ray data of HD\,3191 do not provide evidence for an X-ray emission typical of an HMXB. 

\section{Discussion \label{sect:discussion}}
\subsection{Summary of observational properties of HD\,3191}
The photometric time series revealed the presence of multiperiodic variations with two dominant frequencies $\nu_1 = 0.1235$\,d$^{-1}$ and $\nu_2 = 1.6038$\,d$^{-1}$ with amplitudes near 12\,mmag, as well as several additional frequencies producing modulations at the 1 - 2\,mmag level. Whilst the data suggest temporal changes in the amplitudes, the overall properties of the photometric modulations seem rather stable over the timescale of the {\it TESS} observations.   

\begin{figure}[h]
    \begin{center}
          \resizebox{8.5cm}{!}{\includegraphics{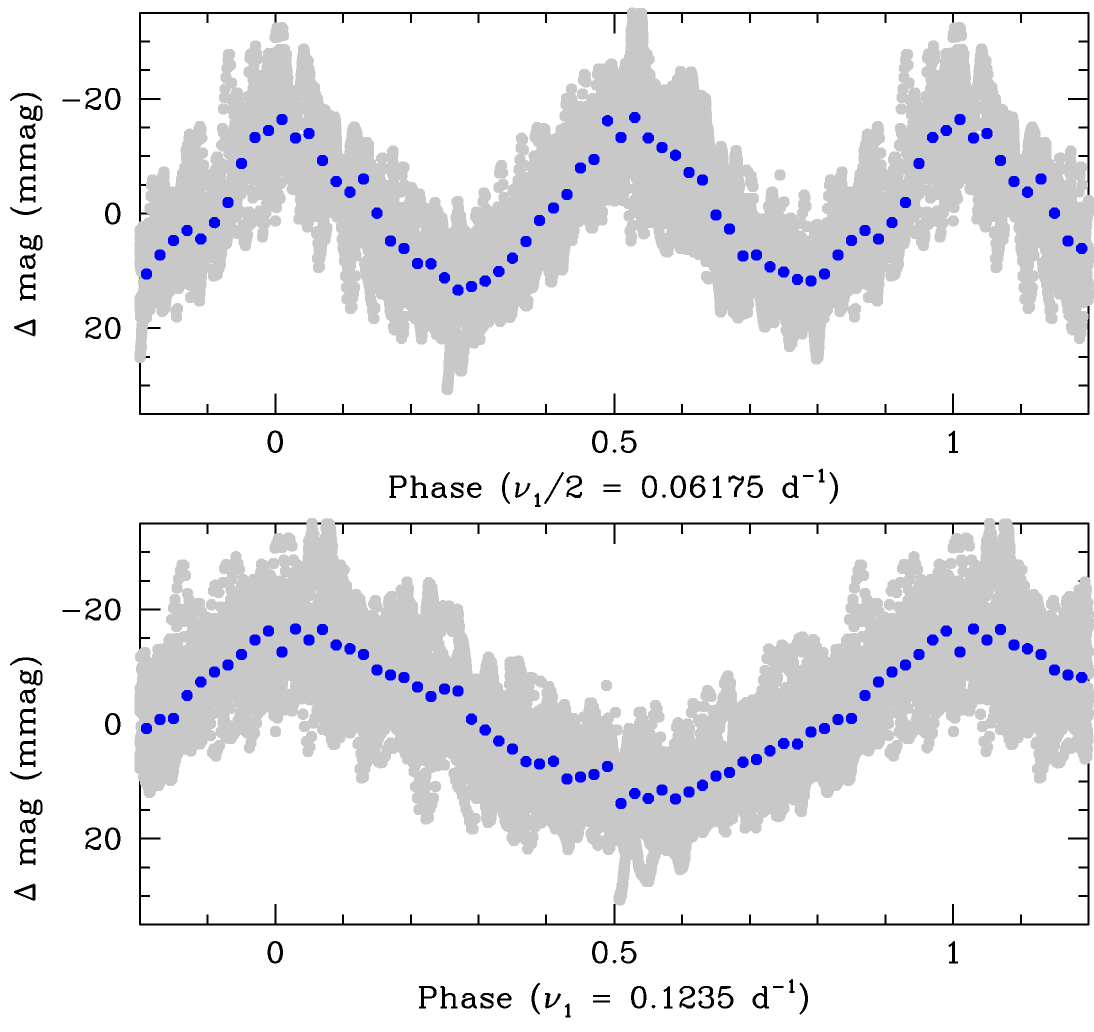}}
          \caption{Residuals of the {\it TESS} photometry after prewhitening all frequencies from Table\,\ref{tab_comb} except for $\nu_1$ (grey symbols). The data are folded in phase with $\nu_1/2$ (top panel) and $\nu_1$ (bottom panel). The blue symbols yield the mean curve, obtained by binning the data in 50 equal phase bins. Phase 0.0 was taken at HJD\,2458762.15 for both plots.\label{orbit_folding}}
    \end{center}
\end{figure}

We reconstructed the variations with the $\nu_1$ frequency by prewhitening the whole {\it TESS} photometry for variations with the frequencies from Table\,\ref{tab_comb} except for $\nu_1$. The resulting residuals were folded with $\nu_1/2$ or $\nu_1$, and binned into 50 equal phase bins. The modulation has a roughly sinusoidal shape with a semi-amplitude of about 18\,mmag (see Fig.\,\ref{orbit_folding}).

Our spectra did not reveal any He\,{\sc ii} $\lambda$~4686 emission nor any clear RV variations on the photometric frequencies. Instead, we found strong line profile variability which hints at the presence of non-radial pulsations. The X-ray data indicate an X-ray luminosity of $L_{\rm X} = (1.54 \pm 0.20) \times 10^{31}$\,erg\,s$^{-1}$, corresponding to $\log{\frac{L_{\rm X}}{L_{\rm bol}}} = -6.93 \pm 0.07$.
\subsection{HD\,3191 as a binary}
\citet{Mar21} and \citet{Mar23} attributed the photometric modulation at the $\nu_1$ frequency to ellipsoidal variations in a HMXB of orbital period 16.09\,d, that is corresponding to $\nu_1/2$. We thus start our discussion by considering this possibility. Ellipsoidal variations are expected to produce a double-wave modulation of the light curve, making the second harmonic of $\nu_{\rm orb}$ the highest peak in the periodogram. For an orbital frequency equal to $\nu_1/2$, Fig.\,\ref{orbit_folding} shows that the double-wave modulation displays minima of nearly equal depths. This explains the lack of a strong signal in the Fourier periodograms at $\nu_1/2$ (Figs.\,\ref{TESSFour} and \ref{FourTESScomb}).    

\begin{figure}[h]
    \begin{center}
          \resizebox{8.5cm}{!}{\includegraphics{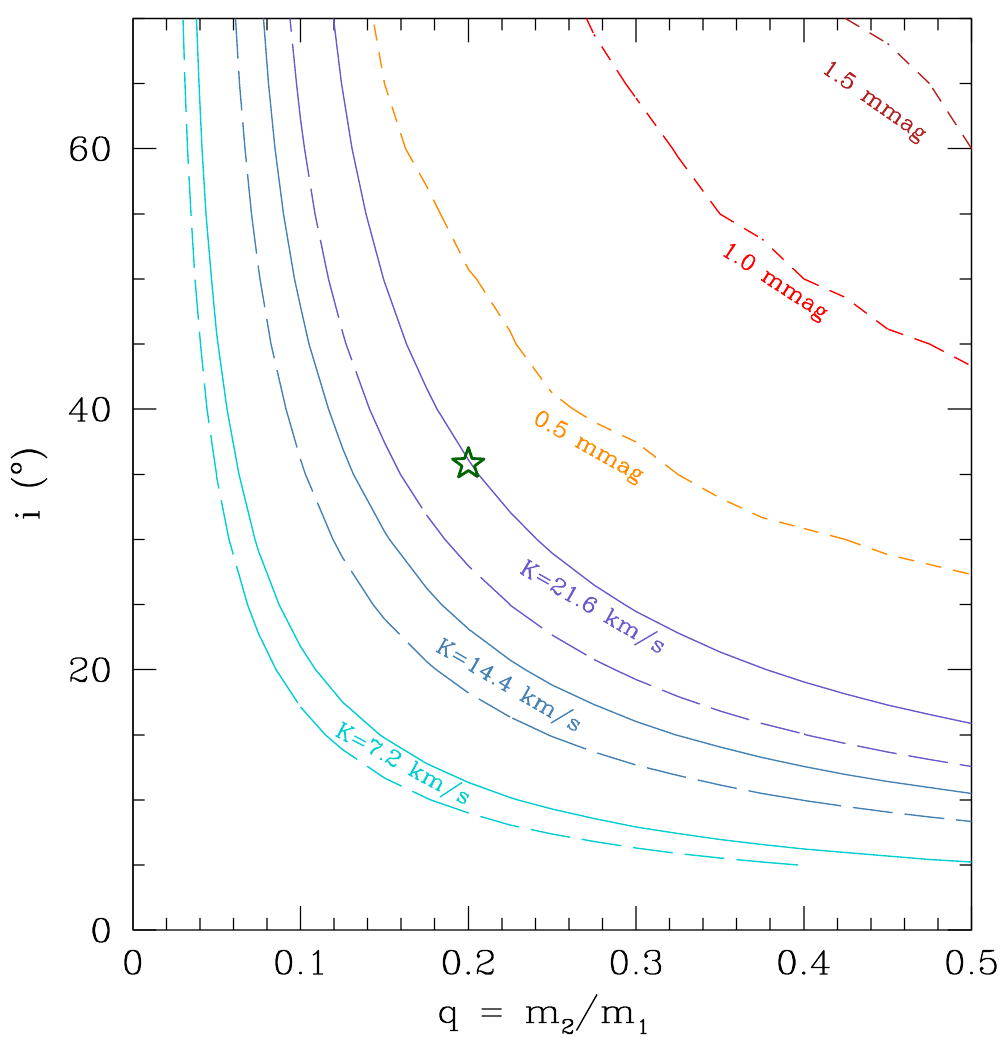}}
          \caption{Contours of same RV semi-amplitude and curves of same photometric peak-to-peak amplitude as a function of mass ratio and orbital inclination.
The calculations assume a binary system hosting a B1\,IV primary of mass 14.8\,M$_{\odot}$, radius 8.34\,R$_{\odot}$ and orbiting on a circular orbit. The solid iso-$K$ curves, and the short-dashed iso-photometric amplitude curves correspond to an orbital period of 16.09\,d. The long-dashed iso-$K$ curves assume a 8.097\,d period. The green star yields the configuration advocated by \citet{Mar21}, with a Roche lobe filling factor near unity, implying a radius more than three times larger than the value adopted in our calculations.\label{contouriq}}
    \end{center}
\end{figure}

The RV semi-amplitude, $K$, (in km\,s$^{-1}$) of the primary star of an SB1 system is given by
\begin{equation}
  K = 212.95\,m_1^{1/3}\,P^{-1/3}\,\sin{i}\,\frac{q}{(1+q)^{2/3}}\,\frac{1}{(1-e^2)^{1/2}},
  \label{eqK}
\end{equation}
where $m_1$ is the mass of the primary in M$_{\odot}$, $P$ is the orbital period in d, $i$ is the orbital inclination, $q = \frac{m_2}{m_1}$, and $e$ is the orbital eccentricity. In the following we adopt $m_1 = 14.8$\,M$_{\odot}$, and $P = 16.09$\,d. Since we lack an orbital solution, $e$ is unconstrained. However, for any combination of $i$ and $q$, equation\,\ref{eqK} indicates that taking $e = 0$ yields a lower limit on $K$. Moreover, the light curve folded with $\nu_1/2$ (see Fig.\,\ref{orbit_folding}) does not unveil strong asymmetries that could hint at a significant orbital eccentricity. We thus compare the lower limits for a circular orbit with the observational constraints. The solid lines in Fig.\,\ref{contouriq} illustrate iso-$K$ contours in the $(q,i)$ plane assuming an orbital period of 16.09\,d. The three contours correspond to $K$ values equal to $1\,\sigma$, $2\,\sigma$ and $3\,\sigma$, where $\sigma = 7.2$\,km\,s$^{-1}$ is the dispersion of the combined RV dataset. The observational constraints confine the plausible range of the $(q,i)$ parameter space to low $q$ or low $i$ values.  

Taken alone, the fact that the RV variations of HD\,3191 are small does not necessarily exclude the possibility of the B star having a compact companion. Indeed, \citet{Ara09} discussed the orbital solutions of two $\gamma$-ray binaries, LS\,I~+61\,303 (B0\,Ve, $P_{\rm orb} = 26.5$\,d, $e = 0.54$) and LS\,5039 (ON6.5\,V((f)), $P_{\rm orb} = 3.9$\,d, $e = 0.34$), finding that both systems had low RV amplitudes of about 20\,km\,s$^{-1}$. Moreover, their RVs displayed large scatter about the orbital solutions ($\sigma \sim 7$\,km\,s$^{-1}$).

However, in the case of HD\,3191, there are additional arguments against binarity which come from the amplitude of photometric modulation at the $\nu_1$ frequency. We used the formula of \citet{Egg83}\footnote{We note that the \citet{Egg83} formalism uses a mass ratio $\frac{m_1}{m_2}$ instead of $\frac{m_2}{m_1}$ that we use here.} to evaluate the radius of the B star's Roche lobe as a function of $q$. From there, we evaluated the Roche lobe filling factor corresponding to the stellar radius $R_* = 8.46$\,R$_{\odot}$ established in Sect.\,\ref{SpT}. For $q$ between 0.025 and 0.5, the Roche lobe filling factor ranges between 0.21 and 0.27. We then used the {\it Nightfall} binary star model \citep{Wichmann} to simulate the photometric variations of a hypothetical binary system of 16.09\,d period consisting of a B1\,IV star orbited by a compact companion of various masses and with different orbital inclinations. The short-dashed lines in Fig.\,\ref{contouriq} provide isocontours of the peak-to-peak photometric amplitude. Over the part of the parameter space where the upper limit on $K$ is consistent with our observational constraints, the predicted photometric variations amount to less than 0.5\,mmag. This is clearly at odds with the observed modulation at the $\nu_1$ frequencies, which have peak-to-peak amplitudes that are more than 50 times larger. Hence, there exists no combination of mass ratio and orbital inclination that simultaneously reproduces the observed photometric modulation, the low RV variations, and the likely dimensions of the B1\,IV star.

Of course, one could object that the above result depends on the adopted stellar radius. For instance, \citet{Mar21} proposed that a system with $q = 0.2$ and $i = 35.8^{\circ}$ could reproduce the light curve assuming a 16.09\,d orbital period. However, the corresponding point is shown in Fig.\,\ref{contouriq} by the green star. To fit the observed light curve amplitude then requires a Roche lobe filling factor near 0.88, implying a mean radius of about 30\,R$_{\odot}$. This is significantly larger than the radius we have inferred. Moreover, as shown by Fig.\,\ref{contouriq}, the expected amplitude of RV variations would be nearly at the $3\,\sigma$ upper limit, obtained from our RV study. Another way to illustrate the above dilemma consists in adopting a Roche lobe filling factor of one and to search for combinations of $q$ and $i$ that would account for the photometric variations and simultaneously lead to a low $K$. Such combinations exist for inclinations between $\sim 20^{\circ}$ (for $q \geq 0.225$) and $\sim 35^{\circ}$ (for $q \leq 0.050$). However, such configurations yield light curves with two minima of unequal depths (due to the difference in temperature between the side facing the L$_1$ point and the opposite side), unlike what we observe. Most of all, they again imply stellar radii exceeding 30\,R$_{\odot}$, that is more than three times bigger than the radius estimated from the {\it Gaia} distance. It is very unlikely that the stellar parameters, including the parallax, have errors that would be large enough to explain such a big discrepancy. In view of the above, a double-wave ellipsoidal light curve seems unable to explain the modulations observed with frequency $\nu_1$.

A binary scenario that remains to be tested is a single wave modulation with $\nu_1$ as the orbital frequency (see bottom panel of Fig.\,\ref{orbit_folding}). A nearly sinusoidal single wave orbital light curve is expected via reflection effects in binary systems with a very strong temperature contrast. There are two possible configurations for such a strong reflection effect to arise in a short-period early-type binary. A first possibility is a post-Roche lobe overflow system where the B1\,IV star is the mass gainer paired with a much hotter and smaller stripped star which acted as the mass donor during mass transfer \citep[e.g.,][]{Wan21,Goe23,NazRau}. Alternatively, such a light curve could reflect a nascent binary system, where the B1\,IV star would be the hotter component paired with a low-mass ($q \lesssim 0.15$) pre-main sequence (PMS) star \citep[e.g.,][]{Jer21,Naz23,Pig24,Naz25}. Some nascent binaries, seen under inclinations $i \gtrsim 68^{\circ}$, display narrow grazing eclipses at the minimum and maximum of the sinusoidal wave \citep[e.g.,][]{Naz23}. The light curve of HD\,3191 does not display such eclipses. Therefore, the photometric variations at frequency $\nu_1$ would be entirely due to reflection effects. The amplitude of the reflection effect not only depends on the values of $q$ and $i$, but also on the temperature ratio of the stars and the solid angle occupied by the cooler star (i.e., the reflector) in the sky of the hotter star. The long-dashed contours in Fig.\,\ref{contouriq} illustrate the iso-$K$ contours in the $(q,i)$ plane assuming an orbital period of 8.097\,d. The permitted regions again correspond to low $q$ values or low orbital inclinations.

For a nascent binary configuration, the reflection arises on the illuminated surface of the PMS component. To simulate the photometric variability of such a putative nascent binary, we used the {\it Nightfall} code assuming $M_2  = 1.2$\,M$_{\odot}$ ($q = 0.08$), $T_{\rm eff, 1} = 27\,000$\,K, and $T_{\rm eff, 2} = 6\,000$\,K. The primary Roche lobe filling factor was set to 0.363 to reproduce a mean radius of 8.46\,R$_{\odot}$. For the secondary Roche lobe filling factor, we took either 0.395 ($R_2 = 3.0$\,R$_{\odot}$) or 0.778 ($R_2 = 6.0$\,R$_{\odot}$). These PMS radii are quite large compared to the values (1 - 4.1\,R$_{\odot}$) found previously in nascent binaries \citep{Naz23,Naz25}, but were adopted here to simulate an optimistic situation with a big reflector surface. The lower value of the secondary filling factor yields maximum peak-to-peak photometric amplitudes of $\sim 14$\,mmag, about half the observed value for $i \leq 74^{\circ}$. Larger inclinations imply eclipses which are not observed. Assuming the larger secondary filling factor produces photometric amplitudes consistent with the observations for $i \simeq 30^{\circ}$. These tests thus show that a nascent binary configuration could possibly account for the constraints on the RVs and the photometric amplitude associated with $\nu_1$, but requires rather large radii and thus a young age (probably $\lesssim 1$\,Myr) for the PMS component. Whilst a PMS companion could also easily explain the observed X-ray properties of HD\,3191, the youth of the PMS star would however be at odds with the evolutionary stage of the B1\,IV star and its 8.8\,Myr age inferred in Sect.\,\ref{SpT}. 

  If the putative companion would be a hot stripped star, the primary B1\,IV star would be the reflector. Hence, the amplitude of the light curve depends on the dimensions of the primary B1\,IV star, for which we found $R_* = 8.46$\,R$_{\odot}$. Assuming $q \simeq 0.1$ ($M_2 = 1.5$\,M$_{\odot}$), $T_{\rm eff, 1} = 27\,000$\,K, and $R_2 = 1.5$\,R$_{\odot}$, we obtain a photometric amplitude 7 -- 8 times lower than observed for $T_{\rm eff, 2} = 50\,000$\,K and $i = 70^{\circ}$. To reproduce the observed photometric amplitude for $i = 70^{\circ}$, we need to assume $T_{\rm eff, 2} = 85\,000$\,K. Lower inclinations obviously require higher temperatures to account for the observed modulation. A temperature $T_{\rm eff, 2} \geq 85\,000$\,K appears quite high in comparison to values of $\leq 45\,000$\,K \citep{Wan21} and $\sim 50\,000$\,K \citep{NazRau} found for stripped stars in B + stripped star systems. However, a stripped star with a temperature of 85\,000\,K was recently reported by \citet{Mue26} as the companion of the Be star MWC\,656.    

We thus conclude that it is very unlikely that HD\,3191 is an HMXB system of period 16.09\,d. In this context, we recall also that the X-ray luminosity $L_{\rm X} = 1.54 \times 10^{31}$\,erg\,s$^{-1}$ is much lower than the X-ray luminosities of HMXBs even considering systems at the lower luminosity end. For instance, the low-luminosity HMXBs LS\,I~+61\,303 and LS\,5039 have $L_{\rm X} \sim 10^{33}$\,erg\,s$^{-1}$ \citep{Sid06,Tak09}. Instead, the X-ray properties fit those usually found for early B-type stars. Nevertheless, we cannot completely exclude the possibility that HD\,3191 could be an 8.097\,d binary with a high temperature ratio. Whilst a stripped star nature for the secondary seems less probable, a nascent binary, with the secondary being a (large) low-mass PMS star, could possibly account for the photometric and RV constraints, but would have trouble accounting for the age of the B1\,IV star. 

\subsection{HD\,3191 as a single star}
We first consider the possibility that $\nu_1$ corresponds to the rotation frequency. Adopting the above inferred stellar radius of $(8.46 \pm 0.76)$\,R$_{\odot}$ then results in an equatorial velocity of $(53.0 \pm 4.8)$\,km\,s$^{-1}$. This value is much lower than the projected equatorial velocity ($v\,\sin{i} = 256 \pm 5$\,km\,s$^{-1}$) evaluated from our spectra, which, by definition, corresponds to a lower limit of the true equatorial velocity. Hence, rotation does not offer a plausible explanation for the $\nu_1$ frequency. Rotation is also unlikely to account for the $\nu_2$ frequency. Indeed, $\nu_2$ exceeds the critical rotation frequency which is about $1.31$\,d$^{-1}$. 

The most obvious alternative to rotational modulation is multi-mode pulsations. Based on the {\it TESS} light curves from Sectors 17 and 18, and their periodograms, \citet{Bal20} classified HD\,3191 as an SPB pulsator. SPBs display non-radial gravity-mode ({\it g}-mode) pulsations of high radial order. These pulsations are driven by the $\kappa$ mechanism in the partial ionisation zone of iron-group elements \citep[][and references therein]{Mig07b,Dzi09}. Traditionally, SPB stars were considered to be lower-mass counterparts of $\beta$\,Cep type pulsators\footnote{The $\beta$\,Cep stars exhibit multiperiodic pulsations mainly due to low-order gravity and pressure modes \citep[e.g.,][]{Sta05}.} with spectral types in the range B3 -- B9 \citep{Bowman}. Theoretical calculations and space-borne photometry revealed considerable overlap between the instability regions of $\beta$~Cep and SPB pulsators, with the latter extending towards earlier B stars \citep[e.g.,][]{Mig07,Bal20}. Table\,\ref{tab_comb}, which includes frequencies above 2.5\,d$^{-1}$, qualifies HD\,3191 as an SPB + $\beta$\,Cep hybrid pulsator. Compared to typical SPB pulsators, the $\nu_1$ frequency of HD\,3191 appears rather low. Indeed, most SPB pulsators have pulsation periods of less than 4\,d \citep[e.g.,][]{Shi23}. An exception is the B0\,II star HD\,260623 (TIC\,55079633) for which \citet{Shi23} found a pulsation period of 9.145\,d. Like HD\,3191, this object also belongs to the hotter SPBs.

\begin{figure}[h]
    \begin{center}
          \resizebox{8.5cm}{!}{\includegraphics{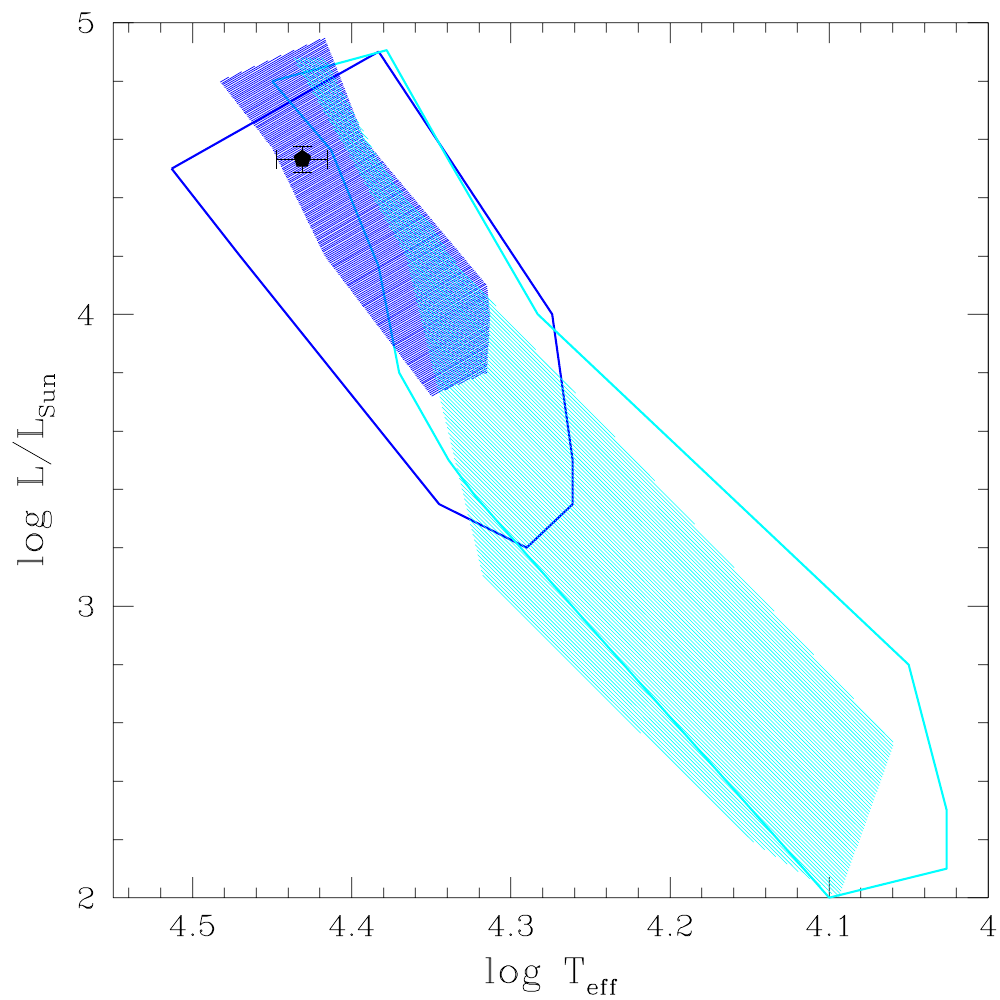}}
          \caption{Location of HD\,3191 (black symbol with error bars) in the $\beta$\,Cep instability strip computed by \citet{Mig07} for either $Z = 0.02$ (blue solid line) or $Z = 0.01$ (blue hatched area) using OP opacities and solar composition according to \citet{Asp05}. The cyan line and hatched area correspond to the SPB instability strip repectively for $Z = 0.02$ and $Z=0.01$ according to the same authors.\label{HRD}}
    \end{center}
\end{figure}

These results concur with the location of the star in the Hertzsprung-Russell diagram. Indeed, adopting the parameters inferred in Sect.\,\ref{SpT} ($T_{\rm eff} = 27\,000$\,K, $L_{\rm bol} = 34040$\,L$_{\odot}$), HD\,3191 is located right in the low-degree ($\ell \leq 3$) $\beta$\,Cep instability strip for $Z = 0.02$ and $Z = 0.1$ as computed by \citet{Mig07}, and lies near the blue boundary of the SPB instability strip computed by the same authors (see Fig.\,\ref{HRD}).

Non-linear resonant coupling between pulsation modes of multiperiodic stars can lead to variability of the frequencies and amplitudes \citep{Buc97,Pig08,Deg09}. A similar process could be at work in HD\,3191, thereby explaining the differences found in our Fourier analysis of the various {\it TESS} sectors.

From our stellar parameters, we find that $\Omega_{\rm rot}/\Omega_{\rm crit} \geq 0.46$. The rapid rotation can affect the properties of the pulsations \citep[e.g.,][and references therein]{Tow04,Tow14}. Non-radial pulsations with $\ell \leq 3$ can be shifted towards lower frequencies by the Coriolis force, which tends to counteract the restoring force on displaced fluid elements \citep{Tow14}. The Coriolis force further distorts the NRPs and traps {\it g} modes into the equatorial region and generates so-called Rossby waves \citep{Tow04}. The latter do not exist in the absence of rotation and their restoring force is the conservation of vorticity. All these effects probably impact the pulsation spectrum of HD\,3191.

\section{Conclusion \label{sect:conclusion}}
Our study has shown that, with the currently available data (optical photometry and spectroscopy, {\it Gaia} distance, X-ray data), HD\,3191 appears very unlikely to be a high-mass X-ray binary. Whilst we cannot rule out the possibility of a nascent binary system, that is an early B-star paired with a low-mass pre-main sequence star with an 8\,d orbital period, the most straightforward explanation for the observational properties appears to be a single star displaying NRPs. Whether the B1\,IV:nn star in HD\,3191 is a genuine single star or is part of a nascent binary, its rapid rotation ($v\,\sin{i} = 256$\,km\,s$^{-1}$) does not seem to be the outcome of a spin-up during a past Roche lobe overflow episode. Similar conclusions have been reached for nascent binaries \citep{Bri24} and for some rapidly rotating OB stars in unevolved wide eccentric binaries \citep{Put18}. Our study thus supports the idea that rapid rotation of massive stars is not always a consequence of past binary interactions. 

\section*{Acknowledgements}
YN acknowledges support from the Fonds National de la Recherche Scientiﬁque (Belgium). PKS was supported by the University of Li\`ege under the Special Funds for Research, IPD-STEMA Programme. This work used data collected by the {\it TESS} mission, which are publicly available from the Mikulski Archive for Space Telescopes (MAST). Funding for the TESS mission is provided by NASA’s Science Mission directorate. ADS and CDS were used during this research.

\appendix
\section{New RV data}
\label{app1}
\begin{figure}[h]
    \begin{center}
          \resizebox{8.5cm}{!}{\includegraphics{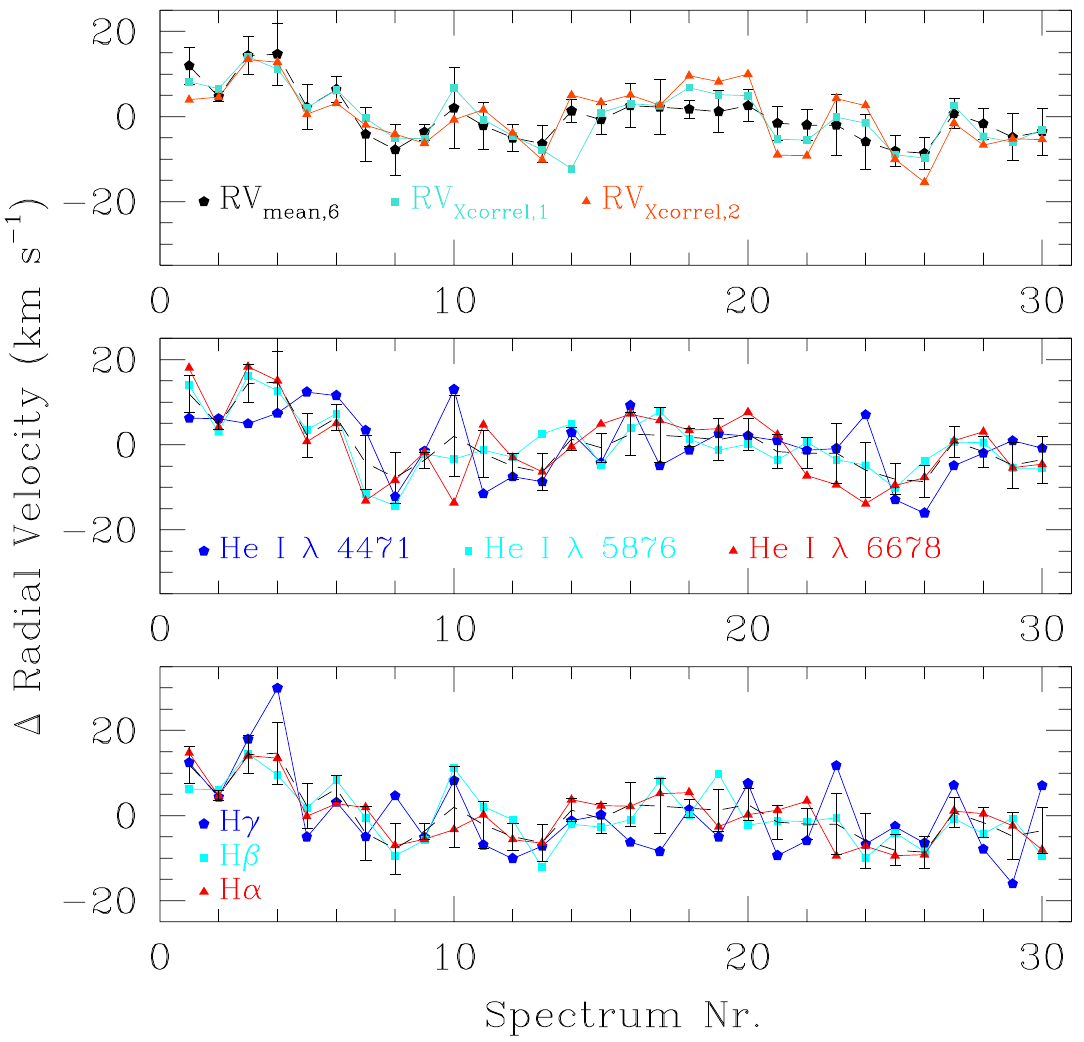}}
      \caption{Bottom and middle panels: RVs of the H\,{\sc i} and He\,{\sc i} lines with the best S/N in our TIGRE/HEROS spectra as a function of the number of the spectrum. The mean RV of each line was subtracted to highlight the relative variations. These mean values are $-32.5$, $-41.4$ and $-43.4$\,km\,s$^{-1}$ respectively for H$\gamma$, H$\beta$ and H$\alpha$, and $-57.5$, $-50.4$ and $-40.5$\,km\,s$^{-1}$ respectively for He {\sc i} $\lambda\lambda$\,4471, 5876 and 6678. The dashed line in each panel yields the mean RV of the six lines (RV$_{\rm mean, 6}(t)$) corrected for its mean value of $-44.3$\,km\,s$^{-1}$. The error bars correspond to the estimated errors on this mean RV. Top panel: comparison between RV$_{\rm mean, 6}(t)$ (black dashed line and pentagons with error bars) and the RVs obtained via cross-correlation with template spectra (see text for details).\label{FigRVspec}}
    \end{center}
\end{figure}

Table\,\ref{newRV} lists the heliocentric dates at mid-exposure of our TIGRE/HEROS spectroscopic observations. We measured the RVs of twelve relatively strong absorption lines in our spectra by fitting a single gaussian to the full line profile. Our measurements include prominent H\,{\sc i} (H9, H$\zeta$, H$\delta$, H$\gamma$, H$\beta$ and H$\alpha$) as well as He\,{\sc i} ($\lambda\lambda$ 4026, 4387, 4471, 4921, 5876, 6678) lines. The results are displayed in Fig.\,\ref{FigRVspec} as a function of the number of the spectrum. We find that different lines exhibit different amplitudes of variations and sometimes different trends (see for instance the RVs of the He\,{\sc i} $\lambda$\,4471 line in Fig.\,\ref{FigRVspec}). Whilst the weakest lines in our sample or those lines with lower S/N ratios (H9, H$\zeta$, He\,{\sc i} $\lambda\lambda$\,4026, 4388, 4921) have dispersions of their RVs up to 16.6\,km\,s$^{-1}$, the strongest lines (H$\alpha$, H$\beta$, He\,{\sc i} $\lambda\lambda$\,4471, 5876) exhibit much lower dispersions between 6.7 and 7.7\,km\,s$^{-1}$.

We computed a mean value RV$_{\rm mean, 12}$ from the RVs of all twelve lines. This quantity exhibits a dispersion of 5.5\,km\,s$^{-1}$. If we restrict the calculation of RV$_{\rm mean, 6}$ to the six strongest lines (H$\gamma$, He\,{\sc i} $\lambda$\,4471, H$\beta$, He\,{\sc i} $\lambda$\,5876, H$\alpha$, He\,{\sc i} $\lambda$\,6678), we obtain a dispersion of 5.9\,km\,s$^{-1}$ (see Table\,\ref{newRV} and Fig.\,\ref{FigRVspec}) which is very similar to the estimated errors on individual RV$_{\rm mean, 6}$ values (between 1.1 and 9.5\,km\,s$^{-1}$ with an average of 4.6\,km\,s$^{-1}$). These latter errors were estimated from the dispersion of the RVs inferred from the six lines. For comparison, the dispersion of the RVs of the narrow (FWHM of 1.2\,\AA) diffuse interstellar band (DIB) at 6613\,\AA\ was found to be of 1.6\,km\,s$^{-1}$, that is much smaller than the errors on the stellar RVs. However, this DIB is much narrower than the stellar lines, allowing a more precise RV determination.

\begin{table}
  \caption{Journal of our TIGRE/HEROS spectroscopic observations. RV$_{\rm mean,6}$ is the mean RV of the six strongest spectral lines (H$\gamma$, He\,{\sc i} $\lambda$\,4471, H$\beta$, He\,{\sc i} $\lambda$\,5876, H$\alpha$, and He\,{\sc i} $\lambda$\,6678), corrected for the mean value of $-44.3$\,km\,s$^{-1}$. The adopted rest wavelengths were 4340.468\,\AA, 4681.332\,\AA, and 6562.850\,\AA\ respectively for the H$\gamma$, H$\beta$ and H$\alpha$ lines, as well as 4471.477\,\AA, 5875.620\,\AA, and 6678.150\,\AA\ for the He\,{\sc i} lines. The last column yields the signal-to-noise ratio in the H$\alpha$ region.\label{newRV}}
  \begin{center}
  \begin{tabular}{c c c c}
    \hline
    Spectrum & HJD$-2\,400\,000$ & RV$_{\rm mean,6}$ & S/N \\
    number & & (km\,s$^{-1}$) & \\
    \hline
 1 & 59495.6765 & $11.9$ & 103 \\
 2 & 59499.7012 & $ 4.7$ & 147 \\
 3 & 59836.8235 & $14.3$ & 119 \\
 4 & 59836.8445 & $14.7$ & 119 \\
 5 & 59864.6623 & $ 2.2$ & 141 \\
 6 & 59864.6833 & $ 6.4$ & 120 \\
 7 & 59869.6564 & $-4.2$ &  86 \\
 8 & 59869.7221 & $-7.8$ & 102 \\
 9 & 59869.7430 & $-3.7$ & 147 \\
10 & 59876.6034 & $ 2.0$ &  75 \\
11 & 59876.6244 & $-2.2$ & 115 \\
12 & 59880.6227 & $-5.0$ & 123 \\
13 & 59880.6436 & $-6.4$ & 161 \\
14 & 60191.8562 & $ 1.3$ & 182 \\
15 & 60191.8771 & $-0.7$ & 152 \\
16 & 60195.7040 & $ 2.5$ & 164 \\
17 & 60199.7556 & $ 2.2$ & 159 \\
18 & 60199.7765 & $ 1.7$ & 130 \\
19 & 60202.6661 & $ 1.2$ & 145 \\
20 & 60202.6870 & $ 2.5$ & 164 \\
21 & 60206.7457 & $-1.6$ & 132 \\
22 & 60206.7667 & $-2.0$ & 139 \\
23 & 60208.7257 & $-2.0$ & 147 \\
24 & 60208.7466 & $-6.0$ & 133 \\
25 & 60210.6693 & $-8.1$ & 122 \\
26 & 60210.6903 & $-8.6$ & 149 \\
27 & 60213.6268 & $ 0.7$ & 185 \\
28 & 60213.6478 & $-1.7$ & 149 \\
29 & 60215.6574 & $-4.9$ & 143 \\
30 & 60215.6783 & $-3.6$ & 156 \\
\hline
  \end{tabular}
  \end{center}
\end{table}

  To check the above results, we further cross-correlated the blue part of the TIGRE/HEROS spectra (between 4200 and 5000\,\AA) with a synthetic TLUSTY spectrum taken from \citet{Lan07}. The template spectrum had an effective temperature 26\,000\,K, $\log{g} = 4.0$ and was rotationally broadened to $v\,\sin{i} = 260$\,km\,s$^{-1}$. The results, corrected for the mean RV of $-31.8$\,km\,s$^{-1}$, are shown by the turquoise symbols in the upper panel of Fig.\,\ref{FigRVspec}. These RVs (named RV$_{\rm Xcorrel, 1}$) have a dispersion around the mean of $6.4$\,km\,s$^{-1}$. With a few exceptions, they agree well with the RV$_{\rm mean, 6}$ values. Finally, since the absorption lines undergo profile variations, we also performed a cross-correlation truncating all the strong lines at a depth of 0.1 times the continuum to avoid their cores. The resulting RVs, corrected for their mean value of $-27.7$\,km\,s$^{-1}$, are displayed as RV$_{\rm Xcorrel, 2}$ in Fig.\,\ref{FigRVspec}. They have a higher dispersion about the mean of $7.1$\,km\,s$^{-1}$, but follow the general trend of the other RV measurements.

\section{Spectral type and rotational velocity}
\label{appSpT}
\begin{figure}[h]
    \begin{center}
          \resizebox{8.5cm}{!}{\includegraphics{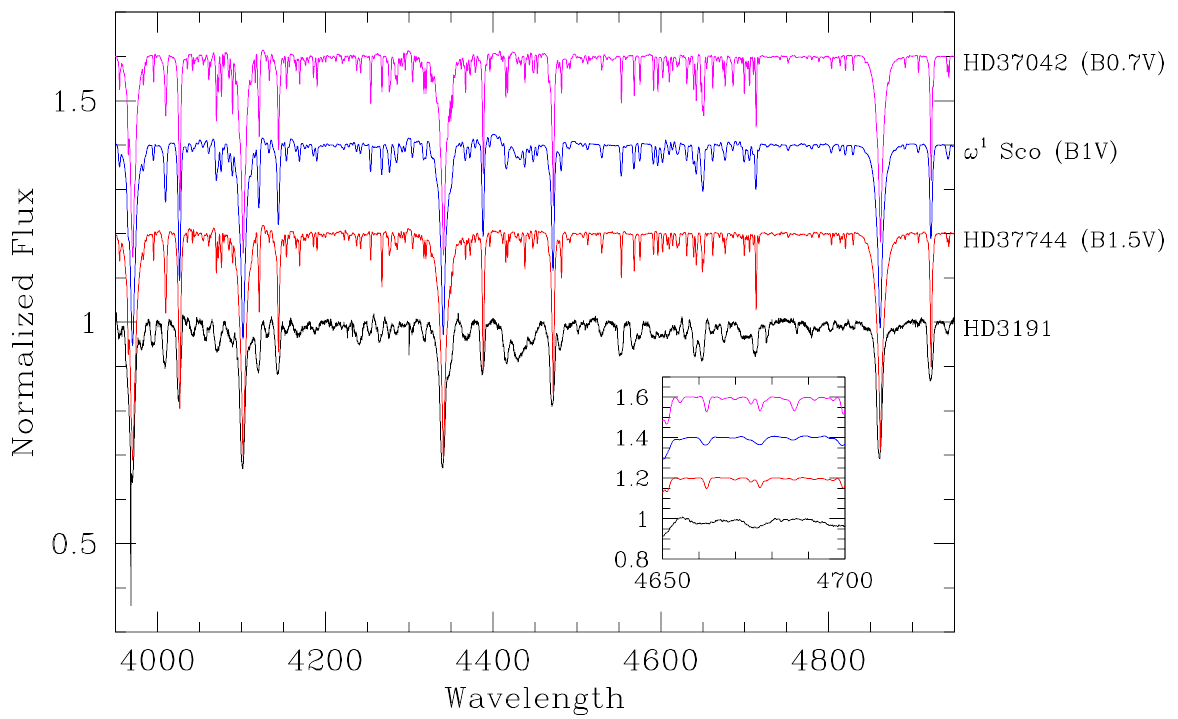}}
          \resizebox{8.5cm}{!}{\includegraphics{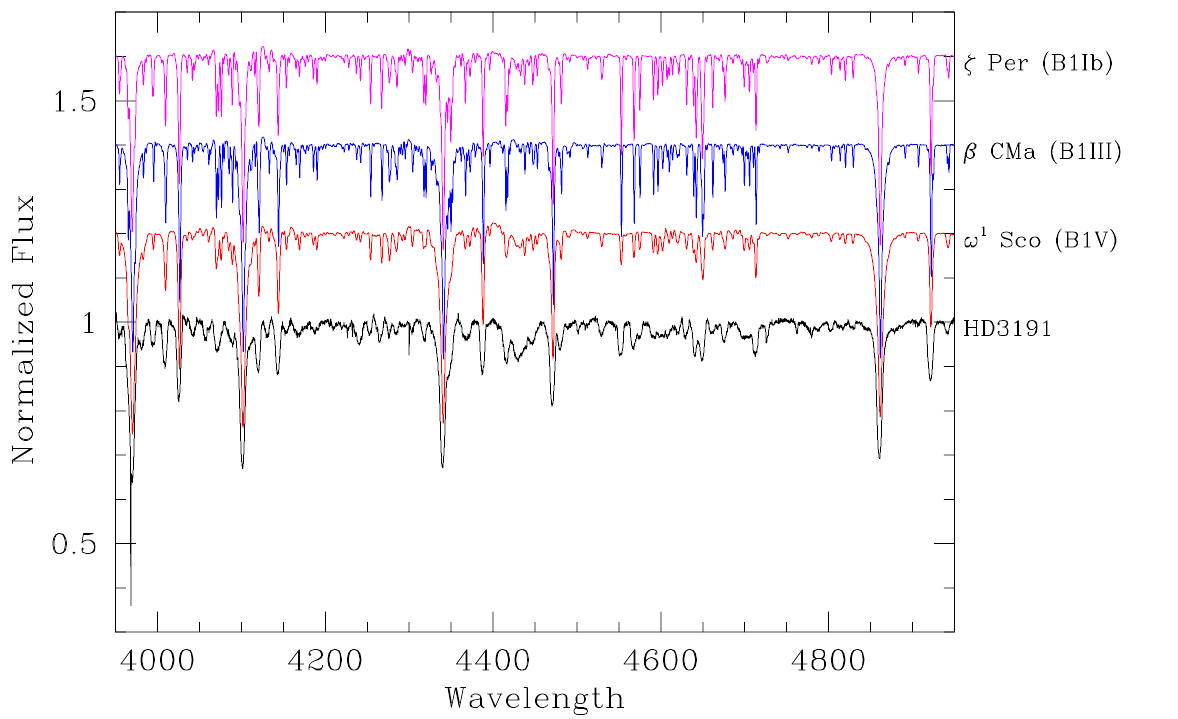}}
      \caption{Comparison of the blue spectrum of HD\,3191 with the spectra of classification standard stars of \citet{Neg24}. The upper panel illustrates the spectra of the main-sequence standard stars HD\,37042 (B0.7\,V), $\omega^1$\,Sco (B1\,V), and HD\,37744 (B1.5\,V), whilst the lower panel provides a comparison with the spectra of the luminosity class standards $\zeta$\,Per (B1\,Ib), $\beta$\,CMa (B1\,III) and $\omega^1$\,Sco. The insert in the upper panel zooms on the 4650 -- 4700\,\AA\ region to emphasise the absence of He\,{\sc ii} $\lambda$\,4686 in the spectrum of HD\,3191.\label{SpT1}}
    \end{center}
\end{figure}

Figure\,\ref{SpT1} illustrates the mean TIGRE/HEROS spectrum compared with the spectra of several classification standard stars for early B-type main-sequence stars taken from the atlas of \citet{Neg24}. Based on the classification criteria of \citet{Wal90} and \citet{Neg24}, the absence of He\,{\sc ii} lines, and more specifically of He\,{\sc ii} $\lambda$\,4686, in the spectrum of HD\,3191 indicates a spectral type later than B0.7. The presence of a weak but definite Si\,{\sc iv} $\lambda$\,4089 line indicates a spectral type earlier than B1.5. Overall, the spectrum of HD\,3191 agrees best with that of the B1\,V standard star $\omega^1$\,Sco, although we note a stronger interstellar absorption as revealed by the strong interstellar Ca\,{\sc ii} absorption blended with H$\epsilon$ and the strong diffuse interstellar bands (DIBs) notably at 4428 and 4726\,\AA. Moreover, the spectral lines of HD\,3191 display a significantly stronger rotational broadening than those of $\omega^1$\,Sco, justifying the nn tag in the spectral classification by \citet{Ree03}. Concerning the luminosity class, while there are no standard stars for luminosity class IV at spectral type B1, we note that comparing the strengths of the Si\,{\sc iii} $\lambda$\,4553 and He\,{\sc i} $\lambda$\,4387 lines indicates a luminosity class V, but certainly no brighter than III. We therefore infer a B1\,V-IV:nn spectral type in good agreement with the spectral classification of \citet{Ree03}.

\begin{figure}[h]
    \begin{center}
          \resizebox{8.5cm}{!}{\includegraphics{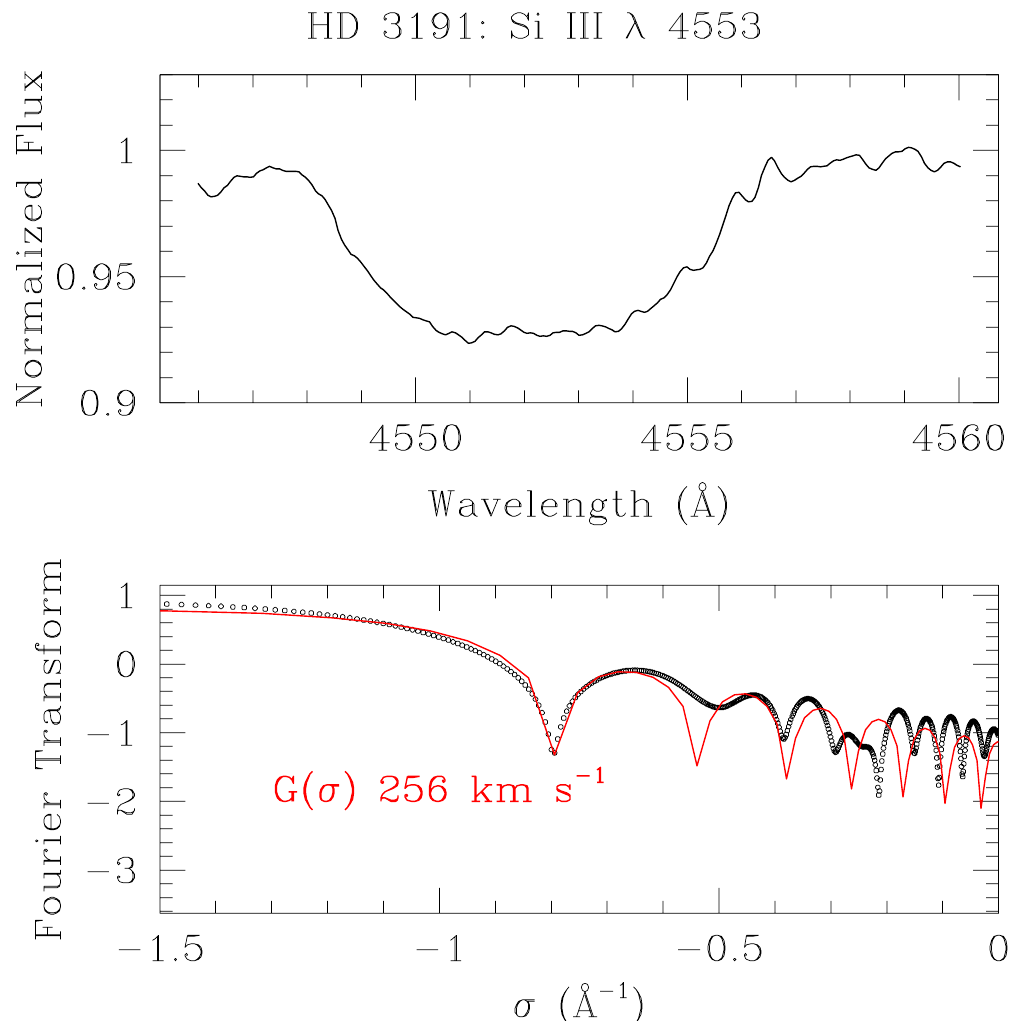}}
          \caption{Determination of the $v\,\sin{i}$ of HD\,3191 via the Fourier transform of the Si\,{\sc iii} $\lambda$\,4553 line. The top panel illustrates the mean line profile , whilst the bottom panel displays the Fourier transform (open dots) along with the Fourier transform of the rotational broadening function for $v\,\sin{i} = 256$\,km\,s$^{-1}$ (red curve).\label{Figvsini}}
    \end{center}
\end{figure}

We used the Fourier method \citep{Sim07} to derive the projected rotational velocity $v\,\sin{i}$ (see Fig.\,\ref{Figvsini}). For this purpose, we selected the Si\,{\sc iii} $\lambda$\,4553 line from our mean TIGRE spectrum. This line is rather isolated and well-suited for deriving the rotational velocity of B1 stars \citep{Sim14}. The best adjustment of the Fourier transform of the line profile was obtained for $v\,\sin{i} = (256 \pm 5)$\,km\,s$^{-1}$, in good agreement with the $(265 \pm 10)$\,km\,s$^{-1}$ value inferred by \citet{Mun16}.  

\section{Extinction towards HD\,3191 and spectral energy distribution}
\label{app2}
      In the compilation of \citet{Ree03}, HD\,3191 has $m_V = 8.58 \pm 0.01$ and $(B-V) = 0.44 \pm 0.01$. \citet{Weg94} quote the intrinsic $(B-V)_0$ of a B1\,V star as $-0.23$ whilst it would be $-0.21$ for a B1\,III star. This implies a colour excess $E(B-V) = 0.66 \pm 0.02$. Assuming $R_V = A_V/E(B-V) = 2.7$ (see hereafter), yields an extinction $A_V = 1.78 \pm 0.06$. Using instead $(B-V)_0 = -0.278$, as given by \citet{Pec13} for B1\,V stars, leads to $E(B-V) = 0.72  \pm 0.01$ and $A_V = 1.94 \pm 0.03$. \citet{Mun16} used the equivalent width of the interstellar K\,{\sc i} $\lambda$ 7696 absorption and the DIB at 6613\,\AA\ to estimate a reddening of $E(B-V) = 0.73$, corresponding to $A_V = 1.97$ (again assuming $R_V = 2.7$). Whilst there exists a strong dispersion in the relationship between reddening and DIB strength, their value agrees reasonably well with ours.
      
\begin{figure}[h]
    \begin{center}
          \resizebox{8.5cm}{!}{\includegraphics{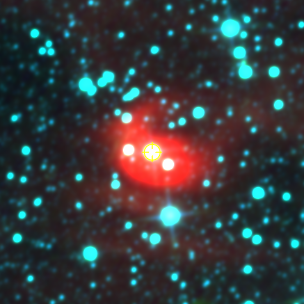}}
          \caption{Three colour near-IR {\it WISE} image of HD\,3191 and its surroundings. The blue, green and red colours correspond respectively to the W1 (3.4\,$\mu$m), W2 (4.6\,$\mu$m), and W4 (22\,$\mu$m) bands. The field of view has a size of 9.9 arcminutes. The position of HD\,3191 is marked with yellow crosshairs.\label{3colWISE}}
    \end{center}
\end{figure}

Using instead the {\tt G-Tomo} tool\footnote{Available via {\tt https://explore-platform.eu/}.} based on the highest resolution {\it Gaia}-2MASS 3D dust maps of \citet{Lal22} and \citet{Ver22}, we obtain $A_V = 1.33^{+.01}_{-.05}$, which corresponds to about 70\% of our estimate. The origin of the discrepant reddening estimates becomes clear by examining images from the {\it Wide-field Infrared Survey Explorer}\footnote{\tt https://irsa.ipac.caltech.edu/data/WISE/docs/release/\\ All-Sky/index.html} \citep[{\it WISE},][]{WISE} all-sky survey. HD\,3191 appears at the centre of a small mid-IR nebula that is seen in the 12\,$\mu$m (W3) and 22\,$\mu$m (W4) bands, but is absent at shorter wavelengths (see Fig.\,\ref{3colWISE}).

The IR nebula has no obvious optical counterpart but is also visible in the 65\,$\mu$m and 90\,$\mu$m band images of the {\it AKARI} all-sky survey \citep{Doi15}. These properties hint at the presence of a large amount of cool dust in the immediate surroundings of HD\,3191, which is most likely responsible for the rather strong extinction towards the star. This dusty structure could be a leftover of the formation process, either of the natal nebula or shed in the course of a stellar merger process. At this stage, we cannot distinguish between both scenarios. However, we note that this blob is fairly compact and does not appear to be directly connected to any surrounding nebular regions.

Owing to the uncertainties on colour calibrations for massive stars, we decided to independently perform a modelling of the spectral energy distribution (SED) using the {\tt speedyfit} Python tool \citep{Vos25}. For this purpose, we first used broadband photometry from {\it Gaia}, {\it 2MASS}, and {\it WISE}. For {\it WISE}, we used only the W1 and W2 data. Indeed, at longer wavelengths HD\,3191 displays a strong IR excess, which is due to the IR nebula described above.

We adopted a Gaussian prior on $T_{\rm eff} = 27\,000$\,K \citep{Nie13} with a dispersion of 1000\,K, a distance prior from \citet{Bai21}, and assumed $R_V = 3.1$. The results ($T_{\rm eff} = (26\,600 \pm 1000)$\,K, $R_* = (8.5 \pm 0.3)$\,R$_{\odot}$, $L_{\rm bol} = (32\,500^{+ 3800}_{- 3500})$\,L$_{\odot}$, $E(B-V) = (0.70 \pm 0.02)$\,mag) are fully consistent with our estimates of Sect.\,\ref{SpT}.

\begin{figure}[h]
    \begin{center}
          \resizebox{8.5cm}{!}{\includegraphics{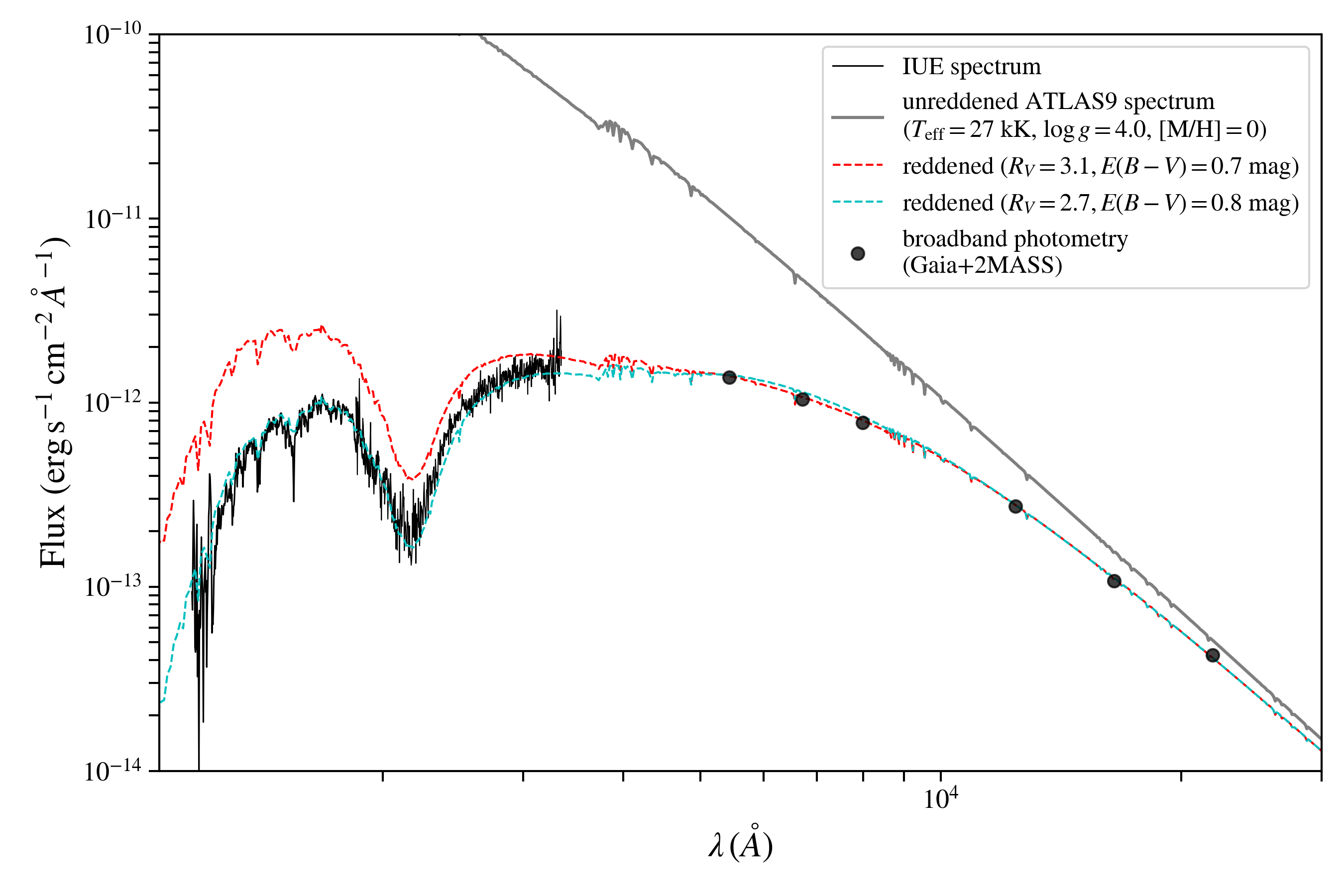}}
          \caption{SED fit to the {\it IUE} spectra and broadband photometry of HD\,3191. The grey continuous line corresponds to an unreddedned Kurucz model atmosphere, whilst the red and cyan dashed lines correspond to the same model reddened respectively with $R_V = 3.1$ or $R_V = 2.7$.\label{SED}}
    \end{center}
\end{figure}

We then compared two flux-calibrated low-dispersion {\it IUE} spectra of HD\,3191 (swp19378 and lwr15413, taken in March 1983), downloaded from MAST, with our SED solution (see Fig.\,\ref{SED}). This comparison clearly reveals that $R_V = 2.7$ and $E(B -V) = 0.8$\,mag yield a much better fit to the SED for the same $T_{\rm eff}$ and $R_*$.
        
The SED of HD\,3191 can be consistently explained by the emission of a B1\,IV star from the far-UV up to the {\it WISE} W2 filter. The {\it IUE} spectra do not display any UV excess that could be indicative of the presence of a relatively bright hotter companion.

\bibliographystyle{elsarticle-harv}
\bibliography{mybiblio}
\end{document}